\documentclass[preprint2]{proto}
\usepackage{times}
\usepackage{color}
\newcommand{\refs}{\par\noindent\hangindent=1pc\hangafter=1}
\newcommand{\me}{$M_{\oplus}$}
\newcommand{\re}{$R_{\oplus}$}

\newcommand{\teff}{$T_{\rm eff}$}

\begin{document}

\title{\textbf{\LARGE Exoplanetary Atmospheres}}

\author {\textbf{\large Nikku Madhusudhan\altaffilmark{1,2}, Heather Knutson\altaffilmark{3}, Jonathan J. Fortney\altaffilmark{4}, Travis Barman\altaffilmark{5,6}}}
\altaffiltext{1}{Yale University, New Haven, CT, USA}
\altaffiltext{2}{University of Cambridge, Cambridge, UK}
\altaffiltext{3}{California Institute of Technology, Pasadena, CA, USA}
\altaffiltext{4}{University of California, Santa Cruz, CA, USA} 
\altaffiltext{5}{Lowell Observatory, Flagstaff, AZ, USA} 
\altaffiltext{6}{The University of Arizona, Tucson, AZ, USA} 

\begin{abstract}
\baselineskip = 11pt
\leftskip = 0.65in 
\rightskip = 0.65in
\parindent=1pc
{\small The study of exoplanetary atmospheres is one of the most exciting and dynamic frontiers in astronomy. Over the past two decades ongoing surveys have revealed an astonishing diversity in the planetary masses, radii, temperatures, orbital parameters, and host stellar properties of exoplanetary systems. We are now moving into an era where we can begin to address fundamental questions concerning the diversity of exoplanetary compositions, atmospheric and interior processes, and formation histories, just as have been pursued for solar system planets over the past century. Exoplanetary atmospheres provide a direct means to address these questions via their observable spectral signatures. In the last decade, and particularly in the last five years, tremendous progress has been made in detecting atmospheric signatures of exoplanets through photometric and spectroscopic methods using a variety of space-borne and/or ground-based observational facilities. These observations are beginning to provide important constraints on a wide gamut of atmospheric properties, including pressure-temperature profiles, chemical compositions, energy circulation, presence of clouds, and non-equilibrium processes. The latest studies are also beginning to connect the inferred chemical compositions to exoplanetary formation conditions. In the present chapter, we review the most recent developments in the area of exoplanetary atmospheres. Our review covers advances in both observations and theory of exoplanetary atmospheres, and spans a broad range of exoplanet types (gas giants, ice giants, and super-Earths) and detection methods (transiting planets, direct imaging, and radial velocity). A number of upcoming planet-finding surveys will focus on detecting exoplanets orbiting nearby bright stars, which are the best targets for detailed atmospheric characterization.  We close with a discussion of the bright prospects for future studies of exoplanetary atmospheres.
 \\~\\~\\~}
\end{abstract}  

\section{\textbf{INTRODUCTION}} 

The study of exoplanetary atmospheres is at the center of the new era of exoplanet science. About 800 confirmed exoplanets, and over 3000 candidates, are now known. The last two decades in exoplanet science have provided exquisite statistics on the census of exoplanets in the solar neighborhood and on their macroscopic properties which include orbital parameters (eccentricities, separations, periods, spin-orbit alignments, multiplicity, etc), bulk parameters (masses, radii,  equilibrium temperatures), and properties of their host stars. The sum total of current knowledge has taught us that exoplanets are extremely diverse in all of these macroscopic physical parameters. We are now entering a new era where we begin to understand the diversity of chemical compositions of exoplanets, their atmospheric processes, internal structures, and formation conditions.  

\begin{figure*}[ht]
\centering
\includegraphics[width = 0.70\textwidth]{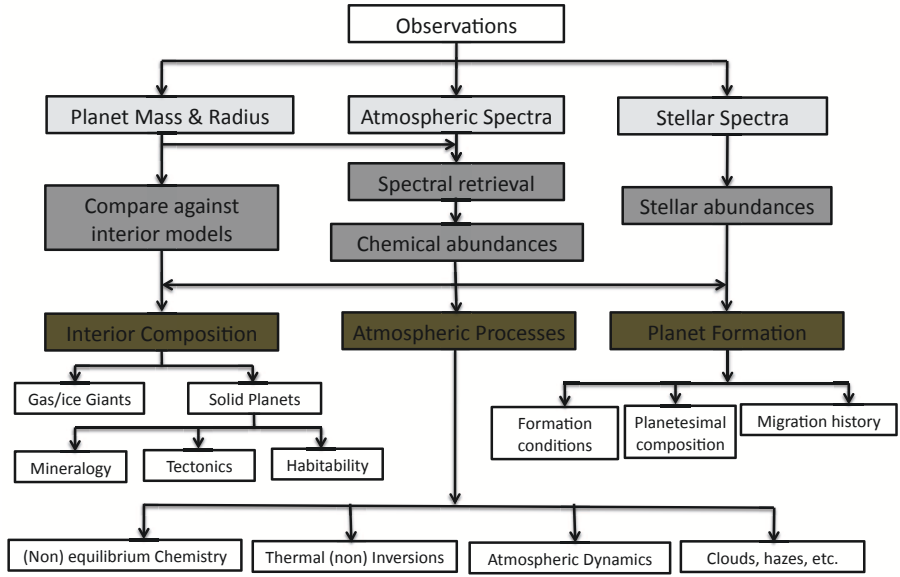}
\caption{Schematic diagram of exoplanet characterization. The top two rows show the observables. The third row shows retrieval methods and/or derived parameters. The remaining rows show the various aspects of exoplanetary atmospheres, interiors, and formation, that can, in principle, be constrained using all the observables.}
\label{fig:schematic}
\end{figure*}

The discovery of transiting exoplanets over a decade ago opened a new door for observing exoplanetary atmospheres. The subsequent years saw tremendous growth in the atmospheric observations of a wide range of transiting exoplanets and also in the detection and characterization of directly imaged planets. Starting with the first detections of atmospheric absorption from a transiting exoplanet (\textit{Charbonneau et al.} 2002) and thermal emission from a transiting exoplanet (\textit{Charbonneau et al.} 2005; \textit{Deming et al.} 2005), atmospheric observations have been reported for more than fifty transiting exoplanets and five directly imaged exoplanets to date. These observations, which are generally either broadband photometry or low resolution spectra, have been obtained using a variety of space-borne and  ground-based observational facilities. Despite the limited spectral resolution of the data and the complexity of the atmospheric models, major advances have been made in our understanding of exoplanetary atmospheres in the past decade. 

When combined with their bulk parameters and host stellar properties, atmospheric spectra of exoplanets can provide simultaneous constraints on their atmospheric and interior properties and their formation histories. Figure 1 shows a schematic overview of exoplanet characterization. The possible observables for an exoplanet depend upon its detection method, but the maximal set of observables, which are possible for transiting exoplanets, comprise of the bulk parameters (mass and radius), atmospheric spectra, and host-stellar spectra. Exoplanetary spectra on their own place constraints on the chemical compositions, temperature profiles, and energy distributions in their atmospheres, which in turn allow constraints on myriad atmospheric processes such as equilibrium/non-equilibrium chemical processes, presence or absence of thermal inversions (`stratospheres' in a terrestrial sense), atmospheric dynamics, aerosols, etc. By combining spectra with the mass and radius of the planet one can begin to constrain the bulk composition of its interior, including the core mass in giant planets and relative fractions of volatiles and rock in rocky planets with implications to interior and surface processes.  Furthermore, the elemental abundances of exoplanetary atmospheres, when compared against those of their host stars, provide insights into their formation conditions, planetesimal compositions, and their evolutionary histories. 

In addition to revealing the diversity of exoplanetary environments, an improved understanding of exoplanetary atmospheres also places the solar system in a cosmic perspective. The exoplanets for which atmospheric characterization is possible now span a large range in mass and effective temperature, including highly irradiated hot Jupiters (T $\sim$ 1300 - 3000 K), warm distant gas giants (T $\sim$ 500-1500 K), hot Neptunes (T $\sim$ 700 - 1200 K), and temperate super-Earths (T $\sim$ 500 K). While the majority of these objects have no direct analogues in the solar system in terms of orbital parameters, the distribution of primary chemical elements (e.g. H, C, O) in their atmospheres can be compared with those of solar system planets. As an important example, the O/H and C/O ratios are poorly known for the solar system giant planets, but could, in principle, be measured more easily for hot Jupiters. Eventually, our knowledge of chemical abundances in exoplanetary atmospheres could enhance our understanding of how planets form and evolve. 

In the present chapter, we review the developments in the area of exoplanetary atmospheres over the past decade. We start with a review in section 2 of the observational methods and available facilities for observing exoplanetary atmospheres. In section 3, we review the theoretical developments which aid in interpreting the observed spectra and in characterizing irradiated exoplanetary atmospheres. In section~4, we review observational inferences for hot Jupiter atmospheres, which are the most observed class of  exoplanets to date. In sections 5-7, we review the theory and observational inferences for other prominent categories of exoplanets: directly-imaged young giant planets, transiting hot Neptunes and super-Earths. We conclude with section 8 where we discuss the future outlook for the field. 

\bigskip
\centerline{\textbf{ 2. OBSERVATIONAL METHODS AND FACILITIES}}
\bigskip

Observational studies of exoplanetary atmospheres can generally be classified into one of two categories: first, measurements of time-varying signals, which include transits, secondary eclipses, and phase curves, and second, spatially resolved imaging and spectroscopy.  These techniques allow us to study the properties of the planet's atmosphere, including the relative abundances of common elements such as C, H, O, and N, corresponding atmospheric chemistries, vertical pressure-temperature profiles, and global circulation patterns. The majority of these observations are made at near-infrared wavelengths, although measurements in the UV, optical, and mid-infrared wavelengths have also been obtained for a subset of planets.  The types of planets available for study with each technique directly reflect the biases of the surveys used to detect these objects.  Transit and radial velocity surveys are most sensitive to massive, short-period planets; these planets also display the largest atmospheric signatures, making them popular targets for time-varying characterization studies.  Spatially resolved imaging is currently limited to young, massive ($\gtrsim$1 $M_{Jup}$) gas giant planets located at large (tens of AU) separations from their host stars, although this will change as a new generation of AO imaging surveys come online.  These planets still retain some of the residual heat from their formation, which makes them hot and bright enough to be detectable at the near-infrared wavelengths of ground-based surveys.     

\begin{figure}[ht]
\centering
\includegraphics[width = 0.4\textwidth]{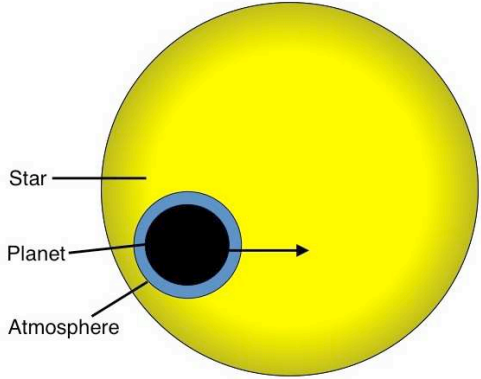}
\caption{Schematic illustration of a planet transiting its host star.  If the planet has an atmosphere (drawn here as a blue annulus) the measured transit depth will vary as a function of wavelength, as the atmosphere will appear opaque at some wavelengths and transparent at others.}
\label{fig:transit}
\end{figure}

In order to interpret these observational data we also need to know the mass and radius of the planet and its host star.  For transiting systems the planet-star radius ratio can be obtained from a simple fit to the transit light curve, and for sufficiently bright stars the planet's mass can be determined from the host star's radial velocity semi-amplitude.  In systems with multiple gravitationally interacting transiting planets, masses can also be estimated from the measured transit timing variations (\emph{Agol et al.} 2005; \emph{Holman and Murray} 2005; \emph{Lithwick et al.} 2012).  For directly imaged planets, masses and radii are typically constrained as part of a global modeling fit to the planet's measured emission spectrum, but the uncertain ages of these systems and the existence of multiple classes of models can result in substantial uncertainties in these fitted parameters.  In some systems dynamical sculpting of a nearby debris disk can also provide an independent constraint on the planet's mass.

In the cases where we have good estimates for the planet's mass and/or radius, we can predict the types of atmospheric compositions that would be consistent with the planet's average density and with our current understanding of planet formation processes.  All of the directly imaged planets detected to date are expected to have hydrogen-dominated atmospheres based on their large estimated masses, while transiting planets span a much wider range of masses and potential compositions.  Although the majority of the transiting planets that have been characterized to date are also hydrogen-dominated gas giants, new discoveries are extending these studies to smaller ``super-Earth"-type planets that may be primarily icy or even rocky in nature.  In the sections below we focus on four commonly used observational techniques for characterizing the atmospheres of extrasolar planets, discussing their advantages and limitations as well as the types of planets that are best-suited for each technique.

\bigskip
\noindent
\textbf{ 2.1 Transmission Spectra}
\bigskip

By measuring the depth of the transit when a planet passes in front of its host star we can determine the size of the planet relative to that of the star. A schematic diagram of a planet in transit is shown in Fig.~\ref{fig:transit}. If that planet has an atmosphere it will appear opaque at some wavelengths and transparent at others, resulting in a wavelength-dependent transit depth.  This wavelength-dependent depth is known as a ``transmission spectrum", because we are observing the absorption features imprinted on starlight transmitted through the planet's atmosphere (\textit{Seager and Sasselov}, 2000; \textit{Brown}, 2001).  This technique is primarily sensitive to the composition of the planet's atmosphere at very low pressures ($\sim~0.1-1000$ mbar) along the day-night terminator, although it is possible to constrain the local pressure and temperature in this region with high signal-to-noise data (e.g., \emph{Huitson et al.} 2012).  Transmission spectroscopy is also sensitive to high-altitude hazes, which can mask atmospheric absorption features by acting as a grey opacity source (e.g., \emph{Fortney} 2005).  The expected depth of the absorption features in a haze-free atmosphere is proportional to the atmospheric scale height:

\begin{equation}
H=\frac{kT}{\mu g}, 
\end{equation}
where, $k$ is Boltzmann's constant, $T$ is the temperature of the planet's atmosphere, $\mu$ is the mean molecular weight of the atmosphere, and $g$ is the surface gravity. The size of the absorbing annulus of the planet's atmosphere can be approximated as $5-10$ scale heights above the nominal planet radius, resulting in a corresponding 
change in the measured transit depth of:
\begin{equation}\label{transm_eq}
\delta_{depth} \simeq \left(\frac{R_p + 10H}{R_*}\right)^2 - \left(\frac{R_p}{R_*}\right)^2
\end{equation}
where $R_p$ and $R_*$ are the planetary and stellar radii. If we take HD 189733b, a typical hot Jupiter, as an example and assume a globally-averaged temperature of 1100 K (\emph{Knutson et al.} 2009a), a surface gravity of 2140 cm s$^{-2}$ (\emph{Bouchy et al.} 2005; {\it Knutson et al.} 2007), and an atmosphere of H$_2$, this would correspond to a scale height of 210 km.  This planet's nominal 2.5\% transit depth would then increase by 0.1\% when observed at wavelengths with strong atmospheric absorption features. At visible wavelengths we expect to see sodium and potassium absorption in hot Jupiter atmosphere, while in the near-infrared this technique is primarily sensitive to H$_2$O, CH$_4$, CO, and CO$_2$ (e.g., \emph{Seager and Sasselov} 2000; \emph{Hubbard et al.} 2001; \emph{Sudarsky et al.} 2003). These observations constrain the planet's atmospheric chemistry and the presence of hazes, while observations of hydrogen Lyman $\alpha$ absorption and ionized metal lines at ultraviolet wavelengths probe the uppermost layers of the planet's atmosphere and provide estimates of the corresponding mass loss rates for close-in hot Jupiters (e.g., \emph{Lammer et al.} 2003; \emph{Murray-Clay et al.} 2009).  

This technique has resulted in the detection of multiple molecular and atomic absorption features (see Section 4.3), although there is an ongoing debate about the validity of features detected using near-IR transmission spectroscopy with the NICMOS instrument on \emph{HST} (e.g., \emph{Swain et al.} 2008b; \emph{Gibson et al.} 2011; \emph{Burke et al.} 2010; \emph{Waldmann et al.} 2013; \emph{Deming et al.} 2013). This debate highlights an ongoing challenge in the field, namely that there is often instrumental or telluric variability in the data on the same time scales (minutes to hours) as the time scales of interest for transmission spectroscopy.  These time-varying signals must be removed from the data, either by fitting with a functional form that provides a good approximation for the effect (e.g., \emph{Knutson et al.} 2007) or by assuming that the time-correlated signals are constant across all wavelengths and calculating a differential transmission spectrum (e.g., \emph{Swain et al.} 2008b; \emph{Deming et al.} 2013).  This naturally leads to debates about whether or not the choice of functional form was appropriate,  whether the noise varies as a function of wavelength, and how these assumptions affect both the inferred planetary transmission spectrum and the corresponding uncertainties on that spectrum.  This situation is further complicated when the planet is orbiting an active star, as occulted spots or a time-varying total flux from the star due to rotational spot modulation can both cause wavelength-dependent variations in the measured transit depth (e.g., \emph{Pont et al.} 2008, 2013; \emph{Sing et al.} 2011).  Errors in the stellar limb-darkening models used for the fits may also affect the shape of the transmission spectrum; this problem is particularly acute for M stars such as GJ 1214 (\emph{Bean et al.} 2010, 2011; \emph{Berta et al.} 2012a).  General practice in these cases is to use the model limb-darkening profiles where they provide a good fit to the data, and otherwise to fit for empirical limb-darkening parameters. 

\bigskip
\noindent
\textbf{ 2.2 Thermal Spectra}
\bigskip

We can characterize the thermal emission spectra of transiting exoplanets by measuring the wavelength-dependent decrease in light when the planet passes behind its host star in an event known as a secondary eclipse. Unlike transmission spectroscopy, which probe the properties of the atmosphere near the day-night terminator, these emission spectra tell us about the global properties of the planet's dayside atmosphere.  They are sensitive to both the dayside composition and the vertical pressure-temperature profile, which determine if the molecular absorption features are seen in absorption or emission. Most secondary eclipse observations are made in the near- and mid-infrared, where the planets are bright and the star is correspondingly faint. It must be noted that secondary eclipse spectra at shorter wavelengths, e.g. in the visible, could also contain contributions due to reflected or scattered light depending on the planetary albedos, as discussed in sections 2.4 and 4.5. For  thermal emission, we can estimate the depth of the secondary eclipse in the Rayleigh-Jeans limit as follows:

\begin{equation} \label{sec_ecl_eq}
depth = \left(\frac{R_p}{R_*}\right)^2 \left(\frac{T_p}{T_*}\right)
\end{equation}

where $R_p$ and $R_*$ are the planetary and stellar radii, and $T_p$ and $T_*$ are their corresponding temperatures.  We can estimate  an equilibrium temperature for the planet if we assume that it radiates uniformly as a blackbody across its entire surface:

\begin{equation}
T_p = \left(\frac{(1-A)L_*}{16\pi \sigma d^2}\right)^{1/4} \simeq T_*\sqrt{\frac{R_*}{2d}}
\end{equation}

where $A$ is the planet's Bond albedo (fraction of incident light reflected across all wavelengths), $\sigma$ is the Stefan-Boltzmann constant, and $d$ is the planet-star distance.  The right-hand expression applies when we assume that the planet's Bond albedo is zero (a reasonable approximation for most hot Jupiters; \emph{Rowe et al.} 2008; \emph{Burrows et al.} 2008; \emph{Cowan and Agol} 2011b; \emph{Kipping and Spiegel} 2011; \emph{Demory et al.} 2011a; \emph{Madhusudhan and Burrows} 2012) and treat the star as a blackbody; more generally, $A$ is a free parameter, leading to some uncertainty in $T_p$. If the planet instead absorbs and re-radiates all of the incident flux on its dayside alone (see Section 2.3), the predicted temperature will increase by a factor of $2^{1/4}$.  

The first secondary eclipse detections were reported in 2005 for the hot Jupiters TrES-1 (\emph{Charbonneau et al.} 2005) and HD 209458b (\emph{Deming et al.} 2005).  More than fifty planets have now been detected in secondary eclipse, of which the majority of observations have come from the \emph{Spitzer Space Telescope} (see \emph{Knutson et al.} 2010 and \emph{Seager and Deming} 2010 for recent reviews) and span wavelengths of $3.6-24~\micron$. There is also a growing body of ground-based near-infrared measurements, in the $\sim$0.8 - 2.1~$\micron$ range (\emph{Sing and L\'opez-Morales} 2009; \emph{Anderson et al.} 2010; \emph{Croll et al.} 2010,2011;  \emph{Gillon et al.} 2012; \textit{Bean et al.}, 2013; \emph{Mancini et al.} 2013;  \emph{Wang et al.} 2013). 

Most of our understanding of the dayside emission spectra of hot Jupiters comes from the combination of broadband \emph{Spitzer} and ground-based secondary eclipse photometry. The challenge with these observations is that we are often in the position of inferring atmospheric properties from just a few broadband measurements; this can lead to degeneracies in our interpretation of the data (\emph{Madhusudhan and Seager} 2009, 2010). One solution is to focus on solar-composition models that assume equilibrium chemistry, but this is only a good solution if these assumptions largely hold for this class of planets (see sections 3.3 and 4.3).  We can test our assumptions using few well-studied benchmark cases, such as HD 189733b (\emph{Charbonneau et al.} 2008) and HD 209458b (\emph{Knutson et al.} 2008).  For these planets we have enough data to resolve most degeneracies in the model fits, while the larger sample of more sparsely sampled planets allows us to fill in the statistical big-picture view as to whether particular atmospheric properties are common or rare.  We can also search for correlations between the planet's atmospheric properties and other properties of the system, such as the incident UV flux received by the planet (\emph{Knutson et al.} 2010), which tells us about the processes that shape planetary atmospheres.

It is also possible to measure emission spectra for non-transiting planets by searching for the RV-shifted signal of the planet's emission in high resolution near-infrared spectra obtained using large ($\sim$10 m) ground-based telescopes (\emph{Crossfield et al.} 2012b; \emph{Brogi et al.} 2013; \emph{de Kok et al.} 2013), although to date this technique has only been applied successfully to a handful of bright, nearby systems.  In order to overcome the low signal-to-noise ratios of individual spectral features, this technique relies on fits using template model atmospheric spectra in order to detect the RV-shifted features in the planet's emission spectrum. This approach is highly complementary to secondary eclipse observations, which are usually obtained at very low spectral resolution.  

\bigskip
\noindent
\textbf{ 2.3 Thermal Phase Curves and Atmospheric Dynamics}
\bigskip

A majority of the transiting exoplanets known to date have short orbital periods where the timescale for orbital synchronization is much shorter than the age of the system.  As a result, we expect these planets to be tidally locked, with permanent day and night sides.  One question we might ask for such planets is what fraction of the flux incident on the dayside is transported to the night side.  If this fraction is large the planet will have a small day-night temperature gradient, whereas if the fraction is small the planet may have large thermal and chemical gradients between the two hemispheres (see \emph{Showman, Menou, and Cho} 2008 for a review).  

\begin{figure}[ht]
\centering
\includegraphics[width = 0.41\textwidth]{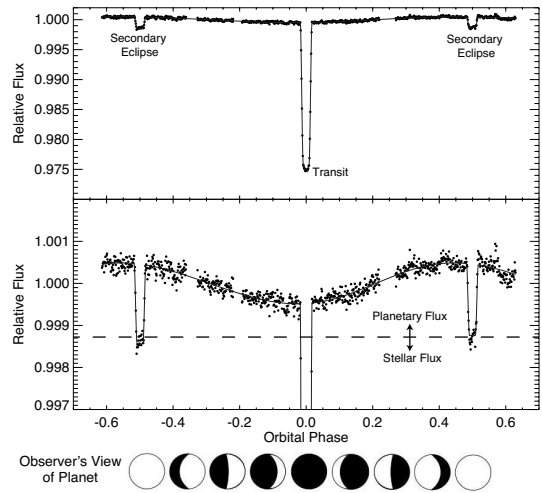}
\caption{Phase curve of hot Jupiter HD~189733b measured in the {\it Spitzer} 4.5 $\mu$m IRAC band.  The transit and secondary eclipses are labeled on the upper panel.  The lower panel shows the same data as the upper panel, but with a reduced y axis range in order to better illustrate the change in flux as a function of orbital phase.  The horizontal dashed line indicates the measured flux level during secondary eclipse, when only the star is visible.  The flux from the planet at any given time can be calculated as the difference between the total measured flux and the dashed line.  Beneath the plot we show a schematic diagram indicating the part of the planet that is visible at different orbital phases, with the planet's day side shown in white and the night side shown in black.}
\label{fig:phase_curve}
\end{figure}

We can estimate the temperature as a function of longitude on these planets and constrain their  atmospheric circulation patterns by measuring the changes in the infrared brightness of the planet as a function of orbital phase (\emph{Cowan and Agol} 2008, 2011a).  
An example of a thermal phase curve is shown in Fig.~\ref{fig:phase_curve}. The amplitude of the flux variations relative to the secondary eclipse tells us the size of the day-night temperature gradient, while the locations of flux maxima and minima indicate the locations of hotter and colder regions in the planet's atmosphere.  If we assume that the hottest region of the planet is near the center of its dayside, then we would expect that the amplitude of the measured thermal phase curve should never exceed the depth of the secondary eclipse.  In the absence of winds we would expect the hottest and coldest regions of the atmosphere to lie at the substellar and anti-stellar points, respectively, corresponding to a flux maximum centered on the secondary eclipse and a flux minimum centered on the transit. Atmospheric circulation models predict that hot Jupiters should develop a super-rotating equatorial band of wind (\emph{Showman and Polvani} 2011); such a wind would shift the locations of these hot and cold regions eastward of their original locations, causing the flux minima and maxima to occur just before the transit and secondary eclipse, respectively. 

If we only have a phase curve observation in a single wavelength, we can treat the planet as a blackbody and invert our phase curve to make a map of atmospheric temperature vs. longitude (\emph{Cowan and Agol} 2008, 2011a).  However, we know that planetary emission spectra are not blackbodies, and if we are able to obtain phase curve observations at multiple wavelengths we can map the planetary emission spectrum as a function of longitude on the planet (\emph{Knutson et al.} 2009a, 2012).  If we assume that there are no compositional gradients in the planet's atmosphere, the wavelength-dependence of the phase curve shape reflects changes in the atmospheric circulation pattern as a function of depth in the atmosphere, with different wavelengths probing different pressures depending on the atmospheric opacity at that wavelength.

There are two additional observational techniques that can be used to place constraints on the atmospheric circulation patterns and wind speeds of transiting planets.  By measuring the wavelength shift of the atmospheric transmission spectrum during ingress, when the leading edge of the planet's atmosphere occults the star, and during egress when the trailing edge occults the star, it is possible to estimate the wind speeds near the dawn and dusk terminators of the planet (\emph{Kempton and Rauscher} 2012; \emph{Showman et al.} 2012).  A marginal detection of this effect has been reported for HD 209458b (\emph{Snellen et al.} 2010b), but it is a challenging measurement requiring both high spectral resolution and a high signal-to-noise ratio for the spectra.  We can also obtain a map of the dayside brightness distributions on these planets by searching for deviations in the shape of the secondary eclipse ingress and egress as compared to the predictions for the occultation of a uniform disk (\emph{Williams et al.} 2006; \emph{Agol et al.} 2010; \emph{Majeau et al.} 2012; \emph{de Wit et al.} 2012). This effect has been successfully measured for HD 189733b at 8~\micron~and gives a temperature map consistent with that derived from the phase curve observations.  Unlike phase curves, which only probe brightness as a function of longitude, eclipse maps also provide some latitudinal information on the dayside brightness distribution. 

\bigskip
\noindent
\textbf{ 2.4 Reflected Light} 
\bigskip

Measurements of the reflected light from transiting extrasolar planets are extremely challenging, as the visible-light planet-star contrast ratio is much smaller than the equivalent value in the infrared.  The best detections and upper limits on visible-light secondary eclipses to date have all come from space missions, including MOST (\emph{Rowe et al.} 2008), CoRoT (\emph{Alonso et al.} 2009a, b; \emph{Snellen et al.} 2009; \emph{Snellen et al.} 2010a), and Kepler (\emph{Borucki et al.} 2009, \emph{D\'esert et al.} 2011a,b; \emph{Kipping and Bakos} 2011; \emph{Demory et al.} 2011a, 2013; \emph{Coughlin and L\'opez-Morales} 2012; \emph{Morris et al.} 2013; \emph{Sanchis-Ojeda et al.} 2013; \emph{Esteves et al.} 2013).  Some of these same observations have also detected visible-light phase curves for these planets, which provide information on the planet's albedo as a function of viewing angle.  Many of the planets observed are hot enough to have some detectable thermal emission in the visible-light bands, and it is therefore useful to combine the visible-light eclipse with infrared observations in order to separate out the thermal and reflected-light components (\emph{Christiansen et al.} 2010; \emph{D\'esert et al.} 2011b).  One challenge is that many of the transiting planets with measured visible-light secondary eclipses orbit relatively faint stars, making infrared eclipse measurements challenging for these targets.

\bigskip
\noindent
\textbf{ 2.5 Polarimetry} 
\bigskip

There have been some theoretical investigations of polarized light scattered from hot Jupiters (\emph{Seager, Whitney, \& Sasselov} 2000, \emph{Stam et al.} 2004, \emph{Madhusudhan \& Burrows} 2012) as well as thermally emitted light from young giant planets (\emph{Marley \& Sengupta} 2011, \emph{de Kok et al.} 2011).  In principle detecting polarized light can constrain the scattering phase function and particle sizes of condensates in a planetary atmosphere, but the data so far for hot Jupiters is unclear. \emph{Berdyugina et al.} (2008) detected a high amplitude polarization signal in B band for the hot Jupiter HD~189733b. However, \emph{Wiktorowicz} (2009) was unable to confirm this result.  Since the information supplied from polarization measurements would be unique and complementary compared to any other observational technique, further observational work is important.

\bigskip
\noindent
\textbf{2.6 Direct Imaging}
\bigskip

Finding low-mass companions (planets or brown dwarfs) to stars by direct
imaging is extremely challenging and many factors come into play. First and
foremost, at the ages of most field stars, even massive planets are
prohibitively faint compared to main-sequence host stars.  Even at the
diffraction limit, planets at separations of less than a few AU will be buried
in the point spread function of the star.  Fortunately, by taking advantage of
planetary evolution and carefully selecting the target stars, direct imaging is
not only possible but provides an exquisite means for atmospheric characterization 
of the planets. 

\begin{figure}[ht]
\centering
\includegraphics[width = 0.41\textwidth]{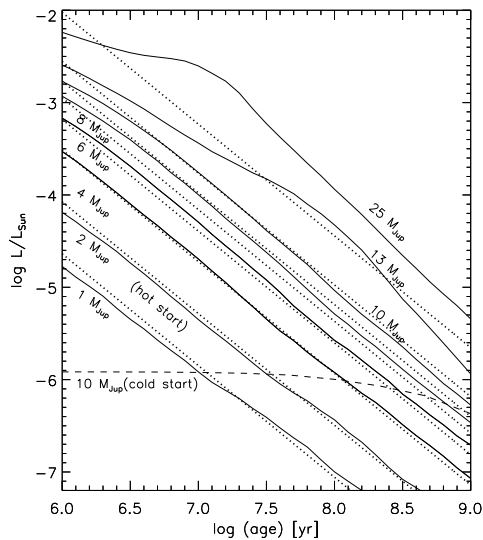}
\caption{Evolution of bolometric luminosity for giant planets.  Solid lines are
hot-start models from \textit{Baraffe et al.} (2003) and dotted lines are based on eq.
\ref{evoleq} for the same masses (see labels). Large changes in  the equation
of state make eq. \ref{evoleq} invalid above $\sim 8$ M$_{\rm jup}$.  The
dashed curve is a cold-start model from \textit{Fortney et al.} (2008b) and illustrates
the impact of different initial conditions.}
\label{fig:evol}
\end{figure}

The bolometric luminosity of gas-giant planets is a smoothly varying function of time, is a
strong function of mass, and is well approximated by 

\begin{equation}
\label{evoleq}
 \frac{L_{\rm bol}(t)}{L_{\odot}} \propto \left(\frac{1}{t}\right)^\alpha M^\beta \kappa^\gamma,
\end{equation}

\noindent where $t$ is time, $M$ is mass and $\kappa$ is the photospheric
Rosseland mean opacity. The exponents $\alpha$, $\beta$, and $\gamma$ are
approximately 5/4, 5/2, and 2/5, respectively (\textit{Stevenson}, 1991).  This equation
is derived under several simplifying assumptions for the interior equations of
state and boundary conditions but, for masses below $\sim$ 10 M$_{Jup}$,
closely matches the predictions of more detailed numerical simulations (Fig.
\ref{fig:evol}).

The mean opacity in eq. \ref{evoleq} is also a function of time but has a
smaller impact on the bulk evolution compared to the effect of the
monochromatic opacities on the spectral energy distribution, especially as
photospheric clouds come and go as a planet cools with time.  Moreover, the
weak dependence on $\kappa$ suggests that predictions for L$_{bol}$ are fairly
robust against uncertainties in atmospheric opacities. This simple model for
planet evolution is often referred to as ``hot start" evolution and is very
close to an upper limit on L$_{\rm bol}$.  However, it ignores the initial conditions
established by the formation process. Fig. \ref{fig:evol} illustrates the
evolution of a planet with initial conditions inspired by core accretion
formation (\textit{Fortney et al.}, 2008b), resulting in a much fainter planet.

The ideal stars for direct imaging are those that provide maximum contrast
between planet light and scattered star light.  Equation \ref{evoleq}
immediately reveals a key aspect of giant planet evolution relevant for direct
imaging searches -- young planets are significantly more luminous than older
planets of equal mass.  Furthermore, effective temperature can be approximated
by a similar equation as 5 (albeit with different exponents and a slightly
weaker mass dependence), indicating that the peak of young planet SEDs shifts
to shorter wavelengths with youth. These two characteristics strongly suggest
that the odds of imaging a planet are greatly improved for planets around young
stars. Ideal stars are also nearby (for favorable angular separation) and have
high proper motion (to easily distinguish bound companions from back/foreground
stars).

Even with carefully selected target stars, direct imaging requires extreme
high-contrast observations typically using adaptive optics and large aperture
telescopes on the ground (e.g., at Keck, VLT, Subaru, and Gemini) or
observations from space (e.g., Hubble). In practice, candidates are confirmed
as companions by establishing common proper motion with the host star over
two or more observing seasons.

Generally, direct imaging surveys as yet have had more to say
about limits than revealing large samples of planets. For example, using the
power-law planet distributions of \textit{Cumming et al.} (2008), \textit{Nielsen et al.} (2010)
find that less than 20\% of FGKM-type stars have planets $> 4$ M$_J$ in orbits
between 60 and 180 AU (depending on the mass-luminosity-age models used).
Examples of discoveries are: 2M1207b (\textit{Chauvin et al.}, 2005), Fomalhaut b (\textit{Kalas
et al.}, 2008), Beta Pic b (\textit{Lagrange et al.}, 2010), 1RXS J1609 b (\textit{{Lafreni{\`e}re}}  et
al., 2010) and the quadruple-planetary system HR 8799 b, c, d, and e (\textit{Marois et
al.}, 2008, 2010). There have been several other recent detections 
of directly imaged planetary/substellar objects orbiting young stars, e.g., $\kappa$ And (\textit{Carson et al.}, 2013), GJ 504 (\textit{Kuzuhara et al.}, 2013), HD 106906  (\textit{Bailey et al.}, 2014), but we note that reliable age estimates are needed in order to obtain accurate masses for these objects (e.g., \textit{Hinkley et al.}, 2013). 

\bigskip
\centerline{\textbf{3. THEORY OF IRRADIATED ATMOSPHERES}} 
\bigskip

\bigskip
\noindent
\textbf{3.1 Atmospheric Models} 
\bigskip

In parallel to the observational advances, the last decade has also seen substantial progress in the modeling of exoplanetary atmospheres and interpretation of exoplanetary spectra. Models have been reported for exoplanetary atmospheres over a wide range of, (1) physical conditions, from highly irradiated giant planets (e.g. hot Jupiters and hot Neptunes) and super-Earths that dominate the transiting planet population to young giant planets on wide orbital separations that have been detected by direct imaging, (2) computational complexity, from one-dimensional (1-D) plane-parallel models to three-dimensional (3-D) general circulation models (GCMs), and (3) thermochemical conditions, including models assuming solar-composition in thermochemical equilibrium as well as those with non-solar abundances and non-equilibrium compositions. 

The complexity of models used depend on the nature of data at hand. For low resolution disk-integrated spectra that are typically observed, 1-D models provide adequate means to derive surface-averaged atmospheric temperature and chemical profiles which are generally consistent with those from 3D models. But, when orbital phase curves are observed, 3D models are necessary to accurately model and constrain the atmospheric dynamics.

\medskip
\noindent
\textbf{3.1.1 Self-consistent Equilibrium Models}
\medskip

Traditionally models of exoplanetary atmospheres have been based on equilibrium considerations. Given the planetary properties (incident stellar irradiation, planetary radius, and gravity) and an assumed set of elemental  abundances, equilibrium models compute the emergent stellar spectrum under the assumptions of radiative-convective equilibrium, chemical equilibrium, hydrostatic equilibrium, and local thermodynamic equilibrium (LTE) in a plane-parallel atmosphere (\textit{Seager and Sasselov}, 1998; \textit{Sudarsky et al.}, 2003; \textit{Seager et al.}, 2005; \textit{Barman et al.}, 2005; \textit{Fortney et al.}, 2006; \textit{Burrows et al.}, 2007,2008a). The constraint of radiative-convective equilibrium allows the determination of a pressure-temperature ($P$-$T$) profile consistent with the incident irradiation and the chemical composition (see section~3.2). The constraint of chemical equilibrium allows the determination of atomic and molecular abundances that provide the source of opacity in the atmosphere (see  section~3.3)  and hydrostatic equilibrium relates pressure to radial distance. Typically, solar abundances are assumed in such models. Additional assumptions in the models include a parametric flux from the interior, parameters representing unknown absorbers (e.g. \textit{Burrows et al.}, 2007, 2008), and prescriptions for energy redistribution from the dayside to the nightside (e.g. \textit{Fortney et al.}, 2006; \textit{Burrows et al.}, 2007). Yet other models adopt $P$-$T$ profiles from equilibrium models, or `gray' solutions (e.g. \textit{Hansen}, 2008; \textit{Guillot et al.}, 2010; \textit{Heng et al.}, 2012), but allow the chemical compositions to vary from solar abundances and chemical equilibrium (\textit{Tinetti et al.}, 2007; \textit{Miller-Ricci et al.}, 2009). 

Equilibrium models are expected to accurately represent exoplanetary atmospheres in regimes where atmospheric dynamics and non-equilibrium processes do not significantly influence the temperature structure and chemical composition, respectively. Perhaps mostly readily understood are cases where radiative equilibrium is not expected in the visible atmosphere.  Such a case is likely found in the atmospheres of hot Jupiters, where extreme dayside forcing and advection should strongly alter the temperature structure from radiative equilibrium (e.g. \textit{Seager et al.}, 2005).  When prescribed abundances are used, it can also be difficult to diagnose abundance ratios that are different than assumed, unless a large number of models are run. 

\medskip
\noindent
\textbf{3.1.2 Parametric Models and Atmospheric Retrieval}
\medskip

Recent advances in multi-band photometry and/or spectroscopy have motivated the development of parametric models of exoplanetary atmospheres. Parametric models compute radiative transfer in a plane-parallel atmosphere, with constraints of LTE and hydrostatic equilibrium, similar to equilibrium models, but do not assume radiative-convective equilibrium and chemical equilibrium that are assumed in equilibrium models. Instead, in parametric models the $P$-$T$ profile and chemical abundances are free parameters of the models. For example, in the original models of \textit{Madhusudhan and Seager} (2009) the temperature profile and molecular abundances constitute 10 free parameters; six parameters for the temperature profile and four for the molecular abundances of H$_2$O, CO, CH$_4$, and CO$_2$, and the models are constrained by the requirement of global energy balance. On the other hand, \textit{Lee et al.} (2012) and \textit{Line et al.} (2012) consider `level-by-level' $P$-$T$ profiles where the temperature in each pressure level (of $\sim$ 100 levels) can be treated as a free parameter. Newer models of carbon-rich atmospheres include additional molecules, e.g. HCN and C$_2$H$_2$, as free parameters (\textit{Madhusudhan} 2012). 

Parametric models are used in conjunction with statistical algorithms to simultaneously retrieve the temperature profile and chemical composition of a planetary atmosphere from the observed spectral and/or photometric data. Whereas early retrieval results used grid-based optimization schemes (\textit{Madhusudhan \& Seager} 2009), recent results have utilized more sophisticated statistical methods such as Bayesian Markov chain Monte Carlo methods (\textit{Madhusudhan et al.} 2011; \textit{Benneke and Seager} 2012; Line et al. 2013) and gradient-descent search methods (\textit{Lee et al.}, 2012, 2013; \textit{Line et al.}, 2012).

\begin{figure}[h]
\centering
\includegraphics[height = 0.4\textheight]{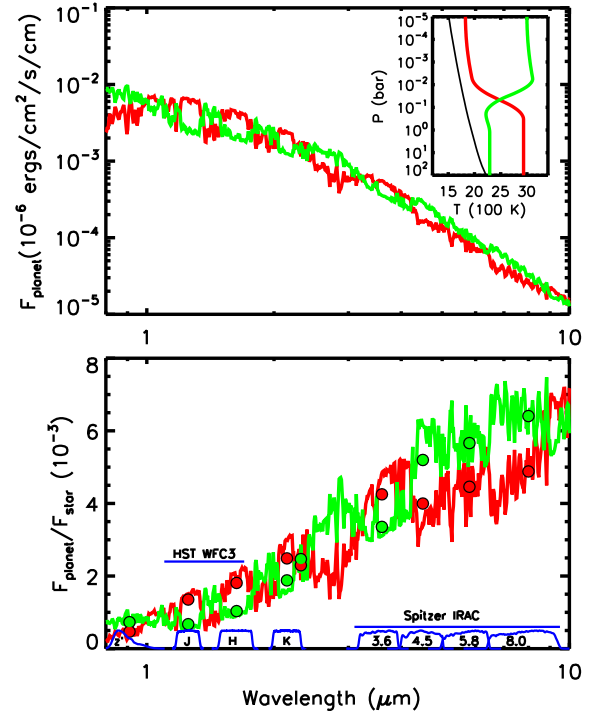}
\caption{Models of dayside thermal emission from a typical hot Jupiter atmosphere (see section 3.1). \emph{Top}:  Emergent spectra for two models with and without a thermal inversion in the atmosphere (corresponding P-T profiles are shown in inset, along with the TiO condensation curve). The model without  (with) a temperature inversion shows molecular bands in absorption (emission). \emph{Bottom}: Spectrum of the planet-to-star flux ratio. The circles show the model binned in the various photometric bandpasses shown.}
\label{fig:thermal_inv}
\end{figure}

\medskip
\noindent
\textbf{3.1.3 Opacities}
\medskip

For any atom, molecule, or ion found in a planetary atmosphere it is important to understand the wavelength dependent opacity as a function of pressure and temperature.  Similarly for materials that condense to form solid or liquid droplets, one needs tabulated optical properties such that cloud opacity can be calculated from Mie scattering theory.  Opacity databases are generally built up from laboratory work and from first-principles quantum mechanical simulation.  The rise of modern high-performance computing has allowed for accurate calculations for many molecules over a wide pressure and temperature range.

These databases have generally also been well-tested against the spectra of brown dwarfs, which span temperatures of $\sim$300-2400 K, temperatures found in hot Jupiters, directly imaged planets, and cooler gas giants as well. Databases for H$_2$O, CO, CO$_2$ and NH$_3$ have recently improved dramatically (e.g. \emph{Rothman et al.},  2010; \emph{Tennyson \& Yurchenko} 2012). A remaining requirement among the dominant molecules is a revised database for methane.  Recent reviews of opacities used in modeling giant planet and brown dwarf atmospheres can be found in \emph{Sharp \& Burrows} (2007) and \emph{Freedman et al.,} (2008). For molecules expected to be common in terrestrial exoplanet atmospheres, there is considerable laboratory work done for Earth's atmospheric conditions.  However, for very hot rocky planets, databases are lacking at higher temperatures.  

\bigskip
\noindent
\textbf{3.2 Theory of Thermal Inversions}
\bigskip

The atmospheric temperature structure involves the balance of the depth-dependent deposition of stellar energy into the atmosphere, the depth dependent cooling of the atmosphere back to space, along with any additional non-radiative energy sources, such as breaking waves.  In terms of radiative transport only, the emergence of thermal inversions can be understood as being controlled by the ratio of opacity at visible wavelengths (which controls the depth to which incident flux penetrates) to the opacity at thermal infrared wavelength, which controls the cooling of the heating planetary atmosphere. Generally, if the optical opacity is high at low pressure, leading to absorption of stellar flux high in the atmosphere, and the corresponding thermal infrared opacity is low, the upper atmosphere will have slower cooling, leading to elevated temperatures at low pressure. Temperature inversions are common in the solar system, due, for instance, to O$_3$ in the Earth, and absorption by photochemical hazes in the giant planets and Titan.

Early models of hot Jupiters did not feature inversions, because early models had no strong optical absorbers at low pressures.  \emph{Hubeny et al.} (2003) pointed out that for the hottest hot Jupiters, TiO and VO gas could lead to absorption of incident flux high in the atmosphere, and drive temperature inversions, as shown in Fig.~\ref{fig:thermal_inv}. This was confirmed by Fortney et al. (2006) and further refined in \emph{Fortney et al. (2008a)} and \emph{Burrows et al.} (2008).  

At this time there is no compelling evidence for TiO/VO gas in the atmospheres of hot Jupiters. \emph{Burrows et al.} (2008) in particular has often modeled atmospheres  with an ad hoc dayside absorber, rather than TiO/VO in particular.  Cloud formation is also a ubiquitous process in planetary atmospheres, and cloud formation may actually remove Ti and V from the gas phase, locking these atoms in Ca-Ti-O bearing condensates below the visible atmosphere.  This could keep even the upper atmospheres of the hottest planets free of TiO/VO gas (\emph{Spiegel et al.} 2009).  Cold night sides could also lead to Ti and V condensing out (\emph{Showman et al.} 2009) and being lost from the day side. It may also be possible that high chromospheric emission from the host star could dissociate inversion-causing compounds in the planetary atmosphere, thereby precluding the formation of thermal inversions (\emph{Knutson et al.} 2010). 

Another possibility is that the chemistry in some atmospheres allows for TiO gas to form in abundance, but in other atmospheres it does not, because there is little available oxygen.  \emph{Madhusudhan et al.} (2011b) showed that the atmospheric C/O ratio can control the abundance of TiO; C/O $\geq 1$ can cause substantial depletion of TiO and VO. 

\begin{figure}[ht]
\centering
\includegraphics[width = 0.45\textwidth]{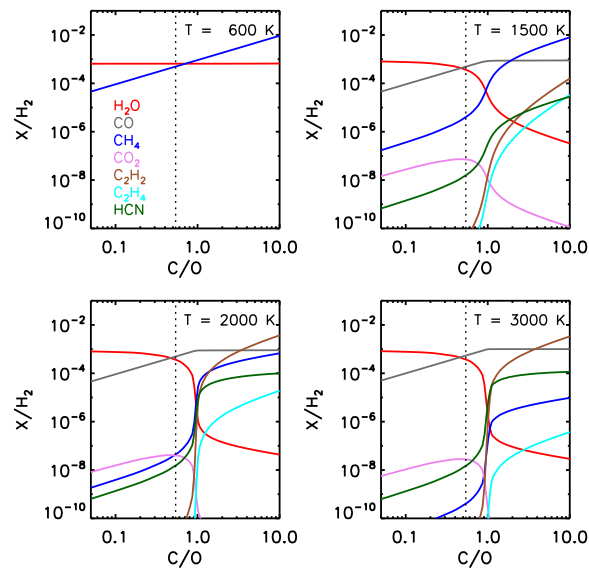}
\caption{Dependance of molecular mixing ratios in chemical equilibrium on C/O ratio and temperature ($T$). Adapted from \emph{Madhusudhan} (2012).} 
\label{fig:cxo_many}
\end{figure}

\bigskip
\noindent
\textbf{3.3 Chemistry in Giant Planetary Atmospheres} 
\bigskip

As discussed in section 3.1, the chemical composition of a planetary atmosphere strongly influences its temperature structure and spectral signatures, and, hence, forms a critical component of atmosphere models.  Extrasolar giant planets, similar to Jupiter and Saturn in our solar system, are expected to host primary atmospheres composed largely of hydrogen (H) and helium (He), as evident from their masses and radii (e.g \textit{Fortney et al.}, 2007; \textit{Seager et al.}, 2007). In addition, the atmospheres are also expected to contain a number of less abundant molecular and atomic species comprising of heavier elements (e.g. O, C, N) in proportions governed by the primordial abundances (\emph{Burrows \& Sharp}, 1999). It is these `trace species' that are responsible for the dominant features of giant planet spectra, and are the primary probes of physico-chemical  processes in their atmospheres and their formation histories. 

\medskip
\noindent
\textbf{3.3.1 Equilibrium Chemistry}
\medskip

It is customary in exoplanet models to assume that the atmospheres are in chemical equilibrium, and to investigate deviations from equilibrium when data necessitates. The chemical composition of a gas in chemical equilibrium at a given temperature ($T$) and pressure ($P$), and elemental abundances, is governed by the partitioning of species that minimizes the Gibbs free energy of the chemical system (e.g. \emph{Burrows \& Sharp}, 1999; \emph{Lodders \& Fegley}, 2002; \textit{Seager et al.}, 2005). Temperatures in atmospheres of currently known irradiated giant exoplanets span a wide range ($\sim$500-3000 K).  For the elemental abundances, it is  customary to adopt solar composition, whereby oxygen is the most dominant element, after H and He, followed by C; the C/O ratio being 0.5 (e.g. \textit{Asplund et al.}, 2009).   

Considering solar abundances and chemical equilibrium, H$_2$O is the dominant carrier of oxygen at all temperatures, whereas carbon is contained in CO and/or CH$_4$ depending on the temperature ($T$); at a nominal pressure of 1 bar, CH$_4$ dominates for T $\lesssim$ 1300 K, whereas CO dominates for higher $T$.  Molecules such as CO$_2$, NH$_3$, and hydrocarbons besides CH$_4$, also become prominent  depending on the temperature (\textit{Burrows and Sharp}, 1999; \textit{Lodders and Fegley}, 2002; \textit{Madhusudhan and Seager} 2011; \textit{Visscher \& Moses}, 2011; \textit{Moses et al.}, 2011). For higher metallicities than the solar, the molar fractions (or `mixing ratios') of all the molecules are systematically enhanced. While solar abundances form a reasonable starting point for atmospheric models, recent studies have shown that the molecular compositions in H-rich atmospheres can be drastically different for different C/O ratios, as shown in Fig.~\ref{fig:cxo_many}. While H$_2$O is abundant in solar composition atmospheres, it can be significantly depleted for C/O $\gtrsim$ 1, where C-rich compounds dominate the composition (\textit{Seager et al.}, 2005; \textit{Helling and Lucas.}, 2009; \textit{Madhusudhan et al.}, 2011b; \textit{Kopparapu et al.}, 2012; \textit{Madhusudhan}, 2012; \textit{Moses et al.}, 2013). 

\medskip
\noindent
\textbf{3.3.2 Non-equilibrium Chemistry}
\medskip

The interplay of various physical processes can drive planetary atmospheres out of chemical equilibrium, resulting in deviations in molecular abundances from those discussed above. Non-equilibrium chemistry in exoplanetary atmospheres have been studied in the context of (a) vertical mixing of species due to turbulent processes on timescales shorter than the equilibrium reaction time scales (also known as `eddy diffusion'), (b) photochemical reactions of species driven by strong incident irradiation, and (c) molecular diffusion or gravitational settling of species. Non-equilibrium chemistry is prevalent in the atmospheres of all solar system giant planets (\textit{Yung and Demore}, 1999) and several brown dwarfs (e.g. \textit{Noll et al.}, 1997), and is similarly expected to be prevalent in the atmospheres of exoplanets (\textit{Zahnle et al.}, 2009; \textit{Moses et al.}, 2011, 2013).

 Several recent studies have reported investigations of non-equilibrium chemistry in exoplanetary atmospheres over a wide range of temperatures, metallicities, and C/O ratios (\textit{Cooper and Showman}, 2006; \textit{Zahnle et al.}, 2009; \textit{Line et al.,} 2010; \textit{Vischer and Moses}, 2010; \textit{Madhusudhan and Seager}, 2011; \textit{Moses et al.}, 2011,2013; \textit{Kopparapu et al.}, 2012; \textit{Venot et al.}, 2012).
A general conclusion of these studies is that non-equilibrium chemistry is most prevalent in cooler atmospheres ($T_{\rm eq} \lesssim 1300$), whereas very hot atmospheres ($T_{\rm eq} \gtrsim 2000$) tend to be in chemical equilibrium in steady state in regions of exoplanetary atmospheres that are accessible to infrared observations, i.e. pressures between $\sim$1 mbar and $\sim$1 bar. 

One of the most commonly detectable effects of non-equilibrium chemistry is the CO-CH$_4$ disequilibrium due to vertical mixing in low-temperature atmospheres. As discussed above, in regions of a planetary atmosphere where $T \lesssim 1300$ K, chemical equilibrium predicts CH$_4$ to be the dominant carbon-bearing species whereas CO is expected to be less abundant. However, vertical fluid motions due to a variety of processes can dredge up CO from the lower hotter regions of the atmosphere into the upper regions where it becomes accessible to spectroscopic observations. Similar disequilibrium can also take place for the N$_2$-NH$_3$ pair in low temperature atmospheres ($T \sim 600-800$ K). 

Models of non-equilibrium chemistry described above span a wide range in complexity, including photochemistry, kinetics, and eddy diffusion, albeit limited to only the major chemical species as necessitated by available data. The most extensive non-equilibrium models reported to date for H$_2$-rich atmospheres still consider reaction networks including only the major elements H, He, O, C, N (\textit{Moses et al.}, 2011,2013), and S (e.g. \textit{Zahnle et al.}, 2009).

\bigskip
\noindent
\textbf{3.4 Atmospheric Dynamics} 
\bigskip

Understanding atmospheric dynamics goes hand in hand with understanding the radiative and chemical properties of atmospheres. For strongly irradiated planets, advection moves energy from day to night and from equator to pole. Additionally, vertical mixing can drive atmospheres away from chemical equilibrium (see section 3.3.2). Several recent studies have reported three-dimensional General Circulation Models (3D GCMs) of exoplanets, particularly for irradiated giant planets for which phase curve data is available to constrain the models (\emph{Cho et al.}, 2008; \emph{Showman et al.}, 2008,2009; \emph{Dobbs-Dixon et al.}, 2010; \emph{Heng et al.}, 2011; \emph{Rauscher and Menou}, 2012). A full review of atmospheric dynamics is beyond the scope of this chapter, but recent reviews for hot Jupiters can be found in \emph{Showman et al.} (2008), for exoplanets in general in \emph{Showman et al.} (2009), and for terrestrial exoplanets in \emph{Showman et al.} (2013). 

Here, we will focus on a small number of relatively simple parameters which to first order help us understand the dynamics.  The first is the dynamical or advective timescale ($\tau_{\rm adv}$), the characteristic time to move a parcel of gas, 
\begin{equation}
\label{trad3}
\tau_{\rm adv} = \frac{R_{\rm p}}{U},
\end{equation}
where $U$ is the characteristic wind speed and $R_{\rm p}$ the planet radius (\emph{Showman \& Guillot} 2002, \emph{Seager et al.} 2005).

What effect this may have on the dynamical redistribution of energy in a planetary atmosphere can be considered after estimating the radiative timescale, $\tau_{\rm rad}$.  Assuming Newtonian cooling, a temperature disturbance relaxes exponentially toward radiative equilibrium with a characteristic time constant $\tau_{\rm rad}$ (e.g., \emph{Goody \& Yung} 1989).  At photospheric pressures this value can be approximated by
\begin{equation}
\label{trad1}
\tau_{\rm rad} \sim \frac{P}{g} \frac{c_{\rm P}}{4 \sigma T^3},
\end{equation}
where $\sigma$ is the Stefan-Boltzmann constant and $c_{\rm P}$ is the specific heat capacity (\emph{Showman \& Guillot} 2002).

It is the ratio of $\tau_{\rm rad}$ to $\tau_{\rm adv}$ that helps to understand how efficient temperature homogenization will be in the atmosphere.  If $\tau_{\rm rad}$ $<<$ $\tau_{\rm adv}$, then the atmosphere will readily cool before dynamics can move energy.  One would expect a large day-night contrast between the hottest part of the atmosphere at the substellar point and the cooler night side. If $\tau_{\rm rad}$ $<<$ $\tau_{\rm adv}$, then temperature homogenization is quite efficient and one would expect comparable temperatures on the day and night sides. A toy model demonstrating this effect in hot Jupiters can be found in \emph{Cowan \& Agol} (2011a).

For hot Jupiters, which are often assumed to be tidally locked, we are actually in a beneficial position for being able to understand the dynamics observationally.  This is due to the flow length scales being planetary-wide in nature.  Tidally locked hot Jupiters rotate with periods of days, while Jupiter itself rotates in 10 hours.  This significantly slower rotation leads to wider bands of winds ($\sim$1-3 for hot Jupiters, rather than $\sim$20 for Jupiter).  These flow scales can be understood via the characteristic Rhines length and Rossby deformation radius (\emph{Showman et al.} 2009). Both predict that faster rotation will lead to narrower bands and smaller vortices. Furthermore, as discussed in section~2.3, the resulting winds in hot Jupiters cause an eastward shift of the dayside hot spot away from the substellar point.

The study of the atmospheric dynamics of young self-luminous planets (such as Jupiter at a young age) is still in its infancy.  \emph{Showman \& Kaspi} (2013) stress that the faster rotation periods of these planets (which are not tidally locked due to their young ages and large distances from their parent stars) will lead to flow that is rotationally dominated. The forcing of these atmospheres is dominated by the internal heat flux, rather than incident stellar flux.  
The radiative-convective boundary in such planets is essentially coincident with the visible atmosphere, rather than being down at hundreds of bars as in typical hot Jupiters.  Future work will likely aim to better understand time variability of surface fluxes, surface heterogeneity, and vertical mixing.

\bigskip
\centerline{\textbf{4. INFERENCES FOR HOT JUPITERS}}
\bigskip

\bigskip
\noindent
\textbf{4.1 Brightness Maps and Atmospheric Dynamics} 
\bigskip

In order to detect the phase curve of an extrasolar planet we must measure changes in infrared flux with amplitudes less than a few tenths of a percent on time scales of several days.  This is challenging for ground-based telescopes, and the only successful phase curve measurements to date have come from the \emph{Spitzer}, \emph{CoRoT}, and \emph{Kepler Space Telescopes}. The first \emph{Spitzer} phase curve measurements in the published literature were for the non-transiting planets $\upsilon$~And (\emph{Harrington et al.} 2006) and HD 179949 (\emph{Cowan et al.} 2007), but without a radius estimate from the transit and a measurement of the dayside flux from the secondary eclipse depth these sparsely sampled phase curves are difficult to interpret.  \emph{Knutson et al.} (2007) reported the first continuous phase curve measurement for the transiting hot Jupiter HD~189733b at 8~\micron~using \emph{Spitzer}.  This measurement indicated that the planet had a dayside temperature of $\sim$1200~K and a nightside temperature of $\sim$1000~K, along with a peak flux that occurred just before secondary eclipse indicating that strong super-rotating winds were redistributing energy from the day side to the night side, consistent with predictions of GCM models (\emph{Showman et al.} 2009). 

To date infrared \emph{Spitzer} phase curves have been published for six additional planets, including HD 149206b (\emph{Knutson et al.} 2009b), HD 209458b (\emph{Crossfield et al.} 2012a), HD 80606b (\emph{Laughlin et al.} 2009), HAT-P-2b (\emph{Lewis et al.} 2013), WASP-12b (\emph{Cowan et al.} 2012, and WASP-18b (\emph{Maxted et al.} 2012).  Two of these planets, HD 80606b and HAT-P-2b, have eccentric orbits and phase curve shapes that reflect the time-varying heating of their atmospheres. HD 189733b is currently the only planet with published multi-wavelength phase curve observations (\emph{Knutson et al.} 2009b, 2012), although there are unpublished observations for several additional planets.  Unlike previous single-wavelength observations, the multi-wavelength phase curve data for HD 189733b are not well-matched by current atmospheric circulation models, which may reflect some combination of excess drag at the bottom of the atmosphere and chemistry that is not in equilibrium. 

\emph{Snellen et al.}~(2009) published the first visible-light measurement of a phase curve using \emph{CoRoT} observations of the hot Jupiter CoRoT-1b. This was followed by a much higher signal-to-noise  visible-light phase curve using \emph{Kepler} observations of the hot Jupiter HAT-P-7b. At these visible wavelengths the measured phase curve is the sum of the thermal and reflected light components, and infrared observations are generally required to distinguish between the two. Most recently, \emph{Demory et al.} (2013) reported a \emph{Kepler} phase curve of the hot Jupiter Kepler-7b and inferred a high geometric albedo  (0.35 $\pm$ 0.02) and the presence of spatially inhomogeneous clouds in the planetary atmosphere. 

We can compare the statistics of atmospheric circulation patterns over the current sample of hot Jupiters.  Although our sample of phase curves is limited, it is possible to obtain a rough estimate of the amount of energy that is transported to the planet's night side by fitting the dayside spectrum with a blackbody and then asking how the estimated temperature compares with the range of expected temperatures corresponding to either no recirculation or full recirculation. \emph{Cowan and Agol} (2011b) published an analysis of the existing body of secondary eclipse data and found that there was a large scatter in the inferred amount of recirculation for planets with predicted dayside temperatures cooler than approximately 2300 K, but that planets hotter than this temperature appeared to have consistently weak recirculation of energy between the two hemispheres.  Existing full-orbit phase curve data support this picture, although WASP-12b may be an exception (\emph{Cowan et al.} 2012).

\bigskip
\noindent
\textbf{4.2 Thermal Inversions }
\bigskip

\medskip
\noindent
\textbf{4.2.1 Inferences of Thermal Inversions in Hot Jupiters}
\medskip

The diagnosis of temperature inversions has always been recognized to be a model dependent process.  A clear indication of an inversion would be molecular features seen in emission, rather than absorption, for multiple species across a wide wavelength range. At present low-resolution emission spectra are available for only a handful of hot Jupiters (e.g., \emph{Grillmair et al.,} 2008;  \emph{Swain et al.,} 2008a,2009), however with no definitive evidence of a thermal inversion.  

Progress, instead, had to rely on wide photometric bandpasses, mostly from \emph{Spitzer} IRAC from 3-10 $\mu$m as well as some ground-based measurements, mostly in $K$ band. These bandpasses average over molecular features, making unique identifications difficult. If the atmosphere is assumed to be in chemical and radiative equilibrium it becomes easier to identify planets with inversions (\textit{Burrows et al.}, 2008; \textit{Knutson et al.}, 2008,2009b), but this assumption may not be accurate for many exoplanetary atmospheres. Even if one assumes a solar C to O ratio it is still usually possible to find equally good solutions with and without temperature inversions for the cases where molecular abundances are allowed to vary as free parameters in the fit ( \textit{Madhusudhan and Seager}, 2010). 

It has long been suggested that there should be a connection between the presence of inversions and a very strong absorber in the optical or near UV, such as TiO or VO (\emph{Hubeny et al.} 2003, \emph{Fortney et al.} 2006a, \emph{Burrows et al.} 2008, \emph{Zahnle et al.} 2009).  However, to date the is no compelling observational evidence for such an absorber.  The difficulty occurs for at least two reasons.  The first is that at occultation, the planet-to-star flux ratio is often incredibly small, on the order of $10^{-4} - 10^{-3}$ in the near-infrared, such that obtaining useful data on the dayside emission/reflection of hot Jupiters is technically quite difficult.  A seemingly easier alternative would be to observe optical absorbers in transmission during the transit.  However, here we are hampered by the fact that the transmission observations probe the terminator region, rather than the dayside.  Since the terminator can be appreciably cooler and receive much less incident flux, the chemical composition of the terminator and daysides can be different. Alternatively, it has been suggested that temperature inversions may also arise from non-radiative processes (e.g.  \emph{Menou}, 2012). 

\medskip
\noindent
\textbf{4.2.2 Non-detections of Thermal Inversions and Possible Interpretations}
\medskip

For a number of planets there is no compelling evidence for a temperature inversion. This essentially means there is no heating source at low pressure that cannot have this excess energy effectively radiated to space. This likely also negates the possibility of a high altitude optical absorber, unless this absorber also was opaque in the thermal infrared. 

There have been several discussions of how the ``TiO hypothesis" could be incorrect.  It was originally pointed out by \emph{Hubeny et al.} (2003) that a cold trap could remove TiO gas from the visible atmosphere.  If the planetary \emph{P--T} profile crosses the condensation curve of a Ti-bearing solid at high pressures ($\sim100-1000$) bars in the deep atmosphere, this condensation would effectively remove TiO gas from the upper atmosphere.  \emph{Spiegel et al.} (2009) investigated this scenario in some detail and found that vigorous vertical mixing would be required to ``break'' the cold trap and transport Ti-bearing grains into the visible atmospheres, where they would then have to be vaporized.  They also pointed out that TiO itself is a relatively heavy molecule, and $K_{\rm ZZ}$ values of $10^7$ cm$^2$s$^{-1}$ are required to even keep the molecule aloft at millibar pressures.

\emph{Showman et al.} (2009) discuss the related phenomenon of the night side code trap, which could also be problematic for TiO gas remaining on the day side.  Super-rotation occurs in hot Jupiters, such that gas is transported around the planet from day to night and night to day.  On the cooler night side, Ti could condense into refractory grains, and sediment down, thereby leaving the gas that returns to the day side Ti-free.  One might then need a planet hot enough that everywhere on the day and night sides temperatures are hot enough to avoid Ti condensation.  Recently \emph{Parmentier et al.} (2013) examined 3D simulations with an eye towards understanding if Ti-rich grains could be kept aloft by vertical motion.  They suggest that temperature inversions could be a transient phenomenon depending on whether  a given dayside updraft were rich or poor in Ti-bearing material, that could then vaporize to form TiO. 

\emph{Knutson et al.}, (2010) demonstrated that the presence or absence of thermal inversions in several systems are correlated with the chromospheric activity of their host stars. In particular, there seems to be a modest correlation between high stellar activity and the absence of thermal inversions. This could mean that high stellar UV fluxes destroy the molecules responsible for the formation of temperature inversions (\emph{Knutson et al.} 2010).

\emph{Madhusudhan} (2012) have suggested that the C/O ratio in hot Jupiter atmospheres could play an important role in controlling the presence or absence of atmospheric TiO, and hence the presence or absence of temperature inversions.  In planets with C/O $>$ 1, CO takes up most available oxygen leaving essentially no oxygen for TiO gas.  In this framework, the strongly irradiated planets that are hot enough that TiO gas would be  expected, but observationally lack temperature inversions, would be carbon-rich gas giants.  

\bigskip
\noindent
\textbf{4.3 Molecular Detections and Elemental Abundances} 
\bigskip

Early inferences of molecules in exoplanetary atmospheres were based on few channels of broadband photometry obtained using {\it Spitzer} and {\it HST}. \textit{Deming et al.} (2005) and \textit{Seager et al.} (2005) reported non-detection of H$_2$O in the dayside of hot Jupiter HD~209458b, observed in thermal emission. On the other hand, \textit{Barman} (2007) reported an inference of H$_2$O at the day-night terminator of HD~209458b. \textit{Tinetti et al.} (2007) inferred the presence of H$_2$O at the day-night terminator of another hot Jupiter, HD~189733b, although the observations have been a subject of debate (see e.g. \textit{Beaulieu et al.}, 2008; \textit{D\'esert et al.}, 2009).

Early attempts have also been made to infer molecules using space-based spectroscopy. The {\it HST} NICMOS instrument (1.8 -- 2.3 $\mu$m) was used to report inferences of H$_2$O,  CH$_4$, and/or CO$_2$ in the atmospheres of hot Jupiters HD~189733b,  HD~209458b, and XO-1b (\textit{Swain et al.}, 2008a; \textit{Swain et al.}, 2009a,b; \textit{Tinetti et al.}, 2010). However, there is ongoing debate on the validity of the spectral features derived in these studies (\textit{Gibson et al.}, 2011; \textit{Burke et al.}, 2010; \textit{Crouzet et al.}, 2012), and in the interpretation of the data (\textit{Madhusudhan \& Seager}, 2009; \textit{Fortney et al.}, 2011). \textit{Grillmair et al.} (2008) inferred H$_2$O in the dayside atmosphere of HD~189733b using {\it Spitzer} IRS spectroscopy (also see \textit{Barman}, 2008), but Madhusudhan \& Seager (2009) suggest only an upper-limit on H$_2$O based on a broader exploration of the model space. 

Recent advances in multi-channel photometry and concomitant theoretical efforts are leading to statistical constraints on molecular compositions. Observations of thermal emission from hot Jupiters in four or more channels of {\it Spitzer} photometry, obtained during {\it Spitzer's} cryogenic lifetime, are now known for $\sim$20 exoplanets (e.g. \textit{Charbonneau et al.}, 2008; \textit{Knutson et al.}, 2008, 2012, \textit{Machalek et al.}, 2008; \textit{Stevenson et al.}, 2010; \textit{Campo et al.}, 2011; \textit{Todorov et al.}, 2011; \textit{Anderson et al.}, 2013). Parallel efforts in developing inverse modeling,  or `retrieval methods', are leading to statistical constraints on molecular abundances and temperature profiles from limited photometric/spectroscopic data (see section 3.1). Using this approach, \textit{Madhusudhan and Seager} (2009) reported constraints on H$_2$O, CO, CH$_4$, and CO$_2$ in the atmospheres of HD~198733b and HD~209458b (also see \textit{Madhusudhan and Seager}, 2010,2011; \textit{Lee et al.}, 2012; \textit{Line et al.}, 2012).  

Constraints on molecular compositions of moderately irradiated exoplanets are beginning to indicate departures from chemical equilibrium. As discussed in section 3.3, low temperature atmospheres provide good probes of non-equilibrium chemistry, especially of CO -- CH$_4$ disequilibrium. \textit{Stevenson et al.} (2010) and  \textit{Madhusudhan and Seager} (2011) reported significant non-equilibrium chemistry in the dayside atmosphere of hot Neptune GJ~436b (but see section 6). Non-equilibrium chemistry could also be potentially operating in the atmosphere of hot Jupiter HD~189733b (\textit{Visscher and Moses}, 2011; \textit{Knutson et al.}, 2012). Furthermore, \textit{Swain et al.} (2010) reported a detection of non-Local Thermodynamic Equilibrium (non-LTE) fluorescent CH$_4$ emission from the hot Jupiter HD~189733b (also see \textit{Waldmann et al.} 2012; but cf  \textit{Mandell et al.} 2011; \textit{Birkby et al.} 2013). 

Nominal constraints have also been reported on the elemental abundance ratios in exoplanetary atmospheres. As discussed in section 2, it has now become routinely possible to detect thermal emission from hot Jupiters using ground-based near-infrared facilities. The combination of space-borne and ground-based observations provide a long spectral baseline (0.8 $\mu$m - 10 $\mu$m) facilitating simultaneous constraints on the temperature structure, multiple molecules, and on quantities such as the C/O ratio. Using such a dataset (e.g. \textit{Croll et al.}, 2011; \textit{Campo et al.}, 2011), 
\textit{Madhusudhan et al.} (2011a) reported the first statistical constraint on the C/O ratio in a giant planet atmosphere, inferring a C/O $\geq$ 1 (i.e. carbon-rich) in the dayside atmosphere of hot Jupiter WASP-12b. However, the observations are currently a subject of active debate (see e.g. \textit{Cowan et al.}, 2012; \textit{Crossfield et al.}, 2012c; \textit{F\"{o}hring} et al., 2013). More recently,  \textit{Madhusudhan} (2012) suggested the possibility of both oxygen-rich as well as carbon-rich atmospheres in several other hot Jupiters. 

 Robust constraints on the elemental abundance ratios such as C/O could help constrain the formation conditions of exoplanets. \textit{Lodders} (2004) suggested the possibility of Jupiter forming by accreting planetesimals dominated by tar rather than water-ice (but cf \textit{Mousis et al.}, 2012). Following the inference of C/O $\geq$ 1 in WASP-12b (\textit{Madhusudhan et al.}, 2011a), \textit{\"{O}berg et al} (2011) suggested that C/O ratios in envelopes of giant planets depend on their formation zones with respect to the various ice lines in the protoplanetary disks. Alternately, inherent inhomogeneities in the C/O ratios of the disk itself may also contribute to high C/O ratios in the giant planets (\textit{Madhusudhan et al.}, 2011b). 

We are also beginning to witness early successes in transit spectroscopy using the {\it HST} WFC3 spectrograph and ground-based instruments. \textit{Deming et al.} (2013) reported high S/N noise transmission spectroscopy of hot Jupiters HD~209458b and XO-1b, and a detection of H$_2$O in HD~209458b. WFC3 spectra have been reported recently for several other hot Jupiters (\emph{Gibson et al.} 2012; \textit{Swain et al.}, 2012; \textit{Huitson et al.}, 2013; \textit{Mandell et al.}, 2013). Low resolution near-infrared spectra of hot Jupiters are also being observed from ground (e.g. \textit{Bean et al.} 2013). 

Recent efforts have also demonstrated the feasibility of constraining molecular abundance ratios of atmospheres of  transiting as well as non-transiting hot Jupiters using high-resolution infrared spectroscopy (see section~2.2). The technique has now been successfully used to infer the presence of CO and/or H$_2$O in the non-transiting hot Jupiter $\tau$ Bootis (e.g. \textit{Brogi et al.}, 2012; \textit{Rodler et al.}, 2012) and the transiting hot Jupiter HD~189733b (e.g. \textit{de Kok et al.}, 2013; \textit{Rodler et al.}, 2013); also see \textit{Crossfield et al.}, (2012c). 

\bigskip
\noindent
\textbf{4.4 Atomic Detections } 
\bigskip

 One of the early predictions of models of hot Jupiters was the presence of atomic alkali absorption in their atmospheres. Spectral features of Na and K, with strong resonance line doublets at 589 and 770 nm, respectively, were predicted to be observable in optical transmission spectra of hot Jupiters (\emph{Seager and Sasselov} 2000). The strong pressure broadened lines of these alkalis are well-known in brown dwarf atmospheres at these same \teff\ values.  To date Na has been detected in several planets including HD 209458b (\emph{Charbonneau et al.} 2002), HD 189733b (\emph{Redfield et al.} 2008), WASP-17b (\emph{Wood et al.} 2010), and others, while K has been detected in XO-2b (\emph{Sing et al.} 2011).  It is not yet clear if the lack of many K detections is significant.

Detailed observations of the well studied pressure-broadened wings of the Na absorption feature can allow for additional constraints on the planetary temperature structure.  Since the line cores of the absorption features are formed at low pressure, and the pressure-broadened wings at higher pressure, models of temperature as a function of height can be tuned to yield the temperature-pressure profile at the terminator of the planet (e.g. \emph{Huitson et al.} 2012).

The escaping exosphere that has been observed around several hot Jupiters is another area where atomic detections have been made.  Most clearly, neutral atomic hydrogen has been observed as a large radius cloud beyond the Roche lobe from two well-studied hot Jupiters, HD 209458b and HD 189733b (\emph{Vidal-Madjar et al.}, 2003; \emph{Lecavelier des Etangs et al.}, 2010; \emph{Linsky et al.}, 2010). A recent review of the theory of exospheres and atmospheric escape from hot Jupiters can be found in \emph{Yelle, Lammer, \& Wing-Huen} (2008).  Neutral and ionized metals have been directly observed or inferred in the escaping exospheres of several planets (\emph{Vidal-Madjar et al.} 2004, \emph{Fossati et al.} 2010, \emph{Astudillo-Defru \& Rojo} 2013).

\bigskip
\noindent
\textbf{4.5 Hazes, Clouds, and Albedos} 
\bigskip

In the absence of Rayleigh scattering or scattering from clouds, all incident flux from a parent star will be absorbed. Rayleigh scattering, mainly from gaseous H$_2$, is important in the relatively cloud-free visible atmospheres of Uranus and Neptune.  However, warm planets will have much more abundant molecular gaseous opacity sources than solar-system ice giants.  Thus, the albedos of exoplanets will generally be determined by the relative importance of clouds.

Much of the early theoretical work on giant exoplanet atmospheres focused on albedos and scattered light signatures (\emph{Marley et al.} 1999, \emph{Seager et al.} 2000, \emph{Sudarsky et al.} 2000).  In warm planets above 1000 K, clouds of silicates (such as forsterite, Mg$_2$SiO$_4$ and enstatite, MgSiO$_3$) were found to dramatically increase albedos for what would otherwise be quite dark planets.

Evidence to date shows that indeed most hot Jupiters are quite dark at optical wavelengths.  Geometric albedos have been measured for several hot Jupiters using data from MOST, CoRoT, and especially Kepler. \emph{Rowe et al.} (2008) used MOST data to derive a geometric aledo ($A_{\rm G}$) of $0.038 \pm 0.045$ for HD 209458b. \emph{Snellen et al.} (2010a) measured $A_{\rm G}=0.164 \pm 0.032$ for CoRoT-2b, but noted that most of this flux likely arose from optical thermal emission, not scattered light.  \emph{Barclay et al.} (2012) and \emph{Kipping and Spiegel} (2011) used Kepler data to find $A_{\rm G}=0.0136 \pm 0.0027$ for TrES-2b, the current ``darkest'' planet.  The bulk of the evidence is that clouds do not have a dominant role in the absorption/scattering of stellar light, for this class of planets.  However, there are outliers.  In particular Kepler-7b was found to have $A_{\rm G}=0.35 \pm 0.02$, likely indicating high silicate clouds (\emph{Demory et al.} 2011a, 2013).

It is at the time of transit, when the planet's transmission spectrum is observed, that clouds may play a greater role in altering the stellar flux. It was originally noted by \emph{Seager \& Sasselov} (2000) that clouds could weaken the spectral features in transmission. \emph{Fortney} (2005) suggested that most such features for nearly all hot Jupiters would be weakened due to minor condensates becoming optically thick at the relevant long slant path lengths through the atmosphere.

The first detection of a hot Jupiter atmosphere was the detection of Na atoms in the atmosphere of HD 209458b (\emph{Charbonneau et al.} 2002) but the feature was weaker than expected, perhaps due to cloud material.  Recent work oin the near infrared (\emph{Deming et al.} 2013) finds weakened water vapor features for both HD 209458b and XO-1b.

The poster child for clouds is HD 189733b, which has an optical spectrum indicating Rayleigh scattering  quite high in the planet's atmosphere (\emph{Pont et al.}, 08; \emph{Sing et al.}, 2011;  \emph{Evans et al.}, 2013). \emph{Lecavelier des Etangs} (2008) suggest a cloud dominated by small Rayleigh scattering silicates.

In no planet is there a definitive determination of the composition of the cloud material.  The clouds may originate from condensation of silicates or other refractory ``rocky'' materials, or from photochemical processes.

\bigskip
\centerline{\textbf{5. INFERENCES FOR DIRECTLY IMAGED PLANETS}} 
\bigskip

\noindent
\textbf{5.1 Connections with Brown Dwarfs}
\bigskip

The search for giant planets and brown dwarfs by direct imaging have gone
hand-in-hand over the decades in part because they have overlapping criteria for good stellar targets.  
In fact, the earliest examples of L and T
brown dwarfs, GD165B and Gl229B respectively, were discovered by imaging
surveys of white dwarfs and nearby field stars (\textit{Becklin et al.}, 1988; \textit{Nakajima
et al.}, 1995).  Numerous brown dwarfs are now known, along with many that are
companions to nearby stars.

\begin{figure}[th]
\centering
\includegraphics[width = 0.40\textwidth]{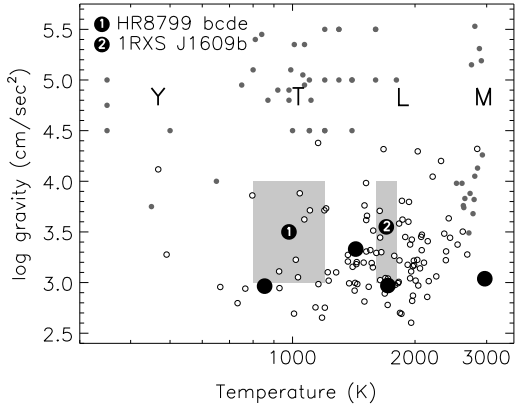}
\caption{
Surface gravities and temperatures for hot-Jupiters (open circles),
brown dwarfs (filled grey circles) and approximate locations for five
planet-mass companions found by direct imaging (gray regions).  Hot-Jupiters
are plotted with equilibrium temperatures assuming 4$\pi$ redistribution and
brown dwarfs are plotted using values from \textit{Rice et al.} (2010), \textit{Burgasser et al.}
(2006), \textit{Cushing et al.} (2011) and \textit{Stephens et al.} (2009). Transiting exoplanets  GJ 436b, HD189733b, HD209458b and WASP-12b are indicated with unnumbered filled symbols, from left to right.}
\label{fig:tefflogg}
\end{figure}

The atmospheres of brown dwarfs, of all ages, provide the best analogues to the
atmospheres of directly imaged planets with the first major similarity being
temperature.  The coldest brown dwarfs currently known have \teff $\sim$ 300K while the youngest, most massive, brown dwarfs have \teff $\sim$ 2500K.  As illustrated in Figure \ref{fig:tefflogg}, this range is expected to
encompass the majority of directly imaged planets.  The masses of directly
imaged planets also overlap with the low-mass end of brown dwarfs, suggesting
similar surface gravities (Fig. \ref{fig:tefflogg}), and convective interiors with thin
radiative H+He envelopes.  Thus, combinations of radiative, convective,
hydrostatic, and chemical equilibrium are likely to be equally as useful
baseline assumptions for the atmospheres of directly imaged planets as they are
for brown dwarfs.  Of course, deviations from equilibrium will occur, just as
they do in some brown dwarfs, the planets in our Solar System and short-period
exoplanets.

Having characteristics that overlap with brown dwarfs allows us to leverage
nearly two decades of observations and modeling that have built detailed 
understanding of the important opacity sources, how temperature and surface
gravity alter spectral energy distributions, and how bulk properties (e.g.
radius) evolve with time and mass.  Brown dwarfs have also been a major impetus
for improvements in molecular line data for prominent carbon, nitrogen and
oxygen-bearing molecules (H$_2$O, CO, CH$_4$, and NH$_3$) and recent years have
seen dramatic improvements (\textit{Tennyson and Yurchenko} 2012).  All these molecules will play an equally important role in understanding directly imaged planets.  Of course,
having significant overlap with brown dwarfs also implies that directly imaged
planets will inherit most of the uncertainty with brown dwarf atmospheres.

\bigskip
\noindent
\textbf{ 5.2 Atmospheric Chemistry}
\bigskip

Observationally, the type of data available for brown dwarfs are also available
for directly imaged planets -- namely traditional spectra and photometry.
Directly imaged planets can be compared, side-by-side, with brown dwarfs in
color-color and color-magnitude diagrams and within sequences of near-IR
spectra. Such comparisons can provide robust identifications of absorption
properties and, in some cases, reliable estimates of temperature and gravity.

Directly imaged planets, because of their youth and masses, will have effective
temperatures comparable to those of many close-in transiting
planets (Fig. \ref{fig:tefflogg}).  However, unlike hot Jupiters, most directly-imaged giant planets
are far enough from their host-star to have minimal stellar irradiation and to not have temperature inversions spanning their near-IR
photospheres. The expectation, therefore, is that, for most of the atmosphere, temperature decreases outward, resulting in deep molecular absorption features comparable to those seen in brown dwarf spectra.

\begin{figure}[ht]
\centering
\includegraphics[width = 0.4\textwidth]{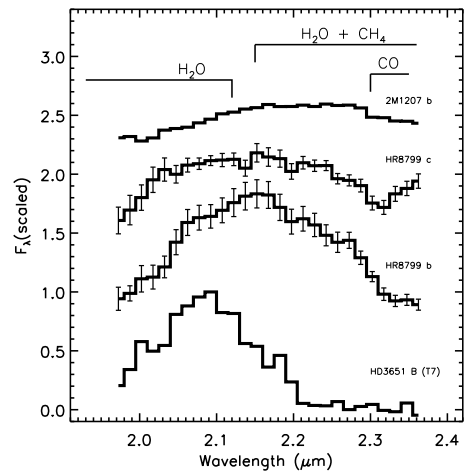}
\caption{
$K$-band spectra for 2M1207b (\textit{Patience et al.} 2012), HR8799c (\textit{Konopacky et al.}
2013), HR8799b (\textit{Barman et al.} 2011a) and HD3651B, all binned to the same
resolution ($R \sim 100$).  Approximate locations of strong absorption bands are
indicated for H$_2$O, CO, and CH$_4$.
}
\label{fig:specseq}
\end{figure}

Figure \ref{fig:specseq} compares low-resolution $K$-band spectra of two young
planets HR8799b and HR8799c (\textit{Barman et al.} 2011; \textit{Konpacky et al.} 2013) to the
young $\sim 5$ M$_{\rm Jup}$ brown dwarf companion 2M1207b (\textit{Patience et al.} 2010) and the old T dwarf HD3651B.  Even at low resolution, water absorption can be
identified as responsible for the slope across the blue side of the $K$ band.
The spectra of 2M1207b and HR8799c have the CO absorption
band-head at $\sim 2.3$ $\mu$m, charactertistic of much hotter brown dwarfs.  Individual CO and H$_2$O lines have been resolved in
spectra of HR8799c, 1RXS J1609, 2M1207b, and many low mass young brown dwarfs.
Methane bands blanket much of the near-IR/IR and, for example, contributes
(along with water) to the very strong absorption seen in spectrum of HD3651b starting
around 2.1 $\mu$m (Fig. \ref{fig:specseq}). So far, methane absorption has not been
unambiguously identified in most young giant planets but likely  contributes to the
slope across the red wing of the $K$-band spectrum of HR8799b (\textit{Barman et al.},
2011; \textit{Bowler et al.},  2010).  Methane absorption is stronger at longer
wavelengths, almost certainly influencing the IR colors across $\sim$ 3
to 5 $\mu$m. Recently, \textit{Janson et al.} (2013) reported a detection of 
methane in the young planet GJ~504b.

As discussed in section 3.3, non-equilibrium chemistry is known to be prevalent in low-temperature planetary atmospheres. Among directly imaged planets and planet-mass companions, HR8799b, HR8799c, and 2M1207b stand out as potentially extreme cases of non-equilibrium chemistry. All three objects have effective temperatures below 1200K but show only CO
absorption and little to no CH$_4$ in their near-IR spectra 
(\textit{Bowler et al.}, 2010; \textit{Barman et al.}, 2011; \textit{Konopacky et al.}, 2013). The HR8799b,c planets are also very cloudy, as inferred from their near-IR broad-band colors (as discussed below),  and the above average cloud cover may also be evidence for strong mixing.
This situation may very well be common in all young low-mass planets with low
surface gravities and suggests that direct imaging  may uncover many cloudy,
CO-rich, planetary atmospheres.

\bigskip
\noindent
\textbf{ 5.3 Cloud Formation}
\bigskip

A major feature of brown dwarf atmospheres is the formation of thick clouds of
dust at photospheric depths beginning around T$_{\rm eff}$ $\sim 2500$K with the
condensation of Fe.  For cooler and less massive field L dwarfs, the impact of
photospheric dust increases as abundant magnesium-silicate condensates are added (e.g. \textit{Helling et al.}, 2008).  
The impact of dust opacity is seen as a steady reddening of the spectral energy distribution 
by about $\sim 0.5$ mags in $H - K$. Mie scattering by small grains dominates the opacity 
for $\lambda < 1$ $\mu$m and is capable of smoothing over all but the strongest absorption 
features.  Field brown dwarfs with T$_{\rm eff}$ $\sim$ 1400K eventually become bluer as 
the location of atmospheric dust shifts to deeper layers and, eventually, the dust is mostly
below the photosphere. Low photospheric dust content is a major characteristic
of T-type brown dwarfs. Eventually, for very low masses and effective
temperatures below $\sim$ 500K, water-ice clouds should form, producing
significant changes in the near-IR colors. 

\begin{figure}[th]
\centering
\includegraphics[width = 0.40\textwidth]{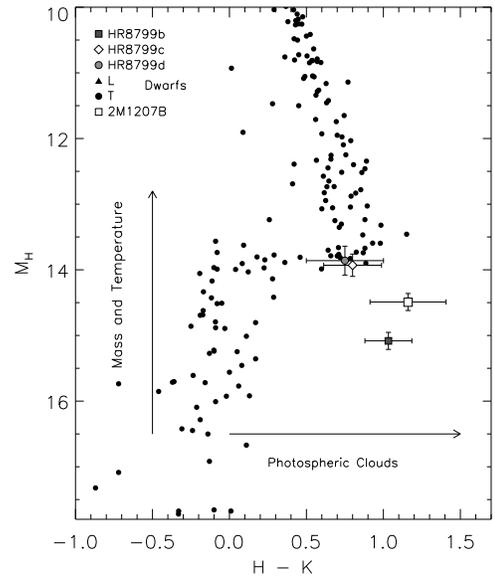}
\caption{
Absolute $H$-band magnitudes versus $H$-$K$ color for field brown dwarfs (\textit{Dupuy and Liu}, 2012) and the young planet-mass companions HR8799bcd and 
2M1207b.  Arrows indicate the basic color and magnitude trends for
increasing photospheric cloud coverage/thickness and mass/temperature, respectively.
}
\label{fig:cmd}
\end{figure}

The reddening of the spectral energy distribution is often measured in
color-color and color-magnitude diagrams (CMD).  Figure \ref{fig:cmd} compares near-IR
colors of field brown dwarfs to several young, directly imaged, companions.
Objects on the reddest part of the CMD ($H-K > 0.5$) have cloudy photospheres
while bluer objects likely having cloud-free or very thin clouds mostly below
the photosphere.   Interestingly the first planet-mass companion found by
direct imaging, 2M1207b, is one of the reddest objects in near-IR CMDs,
indicating that its photosphere is heavily impacted by clouds. The HR8799
planets are also red and, thus, similarly impacted by photospheric dust.
The physical mechanism that causes the photosphere to transition from cloudy to
cloud-free is poorly understood; however, recent observations of young brown
dwarfs suggest that the effective temperature at the transition decreases with
decreasing surface gravity.  Directly imaged planets are not only lower
in mass than brown dwarfs but their surface gravities are also lower because of
their youth. That the first set of directly imaged planets are all
cloudy is further evidence that the transition temperature decreases with
surface gravity and that many young giant planets will have cloudy photospheres
even though their effective temperatures are comparable to the much bluer T
dwarfs.

\bigskip
\noindent
\textbf{5.4 Interpretation of Spectra}
\bigskip

Low to moderate resolution spectroscopy has been obtained for most
planets found so far by direct imaging and discoveries made with the new
instruments, GPI, SPHERE, and P1640, will be accompanied by low resolution
near-IR spectra.  The interpretation of these spectra follows the same
techniques used for brown dwarfs.  Comparisons can be made to empirical
templates spanning the various brown dwarf spectral types to roughly estimate
temperature and gravity. However, this method is unreliable for the
faintest (lowest mass) and reddest objects in Fig. \ref{fig:cmd} because ``normal" brown
dwarfs do not populate the same region of the CMDs. In fact, for 2M1207b, such a
comparison would yield a high effective temperature and unphysical radius
(\textit{Mohanty et al.}, 2007). In the absence of a large set of well-characterized empirical
templates, one is left fitting synthetic model spectra to the observed spectra
to infer temperature, gravity, cloud properties and chemical composition.

When analyzing spectra (or broad photometric coverage) of mature brown dwarfs
(with known distances), it is common to use the measured bolometric luminosity ($L_{\rm bol}$) and evolutionary models to establish a plausible range for both effective
temperature and surface gravity, then restrict model atmosphere fits to within
this range.  This greatly reduces the parameter space one must search
(especially if vertical mixing and cloud properties are important). This method
is fairly robust for old brown dwarfs because the evolution is
well described by ``hot start" models with solar or near-solar abundances (e.g.
\textit{Baraffe et al.}, 2003). However, the initial conditions 
set by the formation process will be important at young ages, making it
difficult to uniquely connect $L_{\rm bol}$ and age to a narrow range of Teff and logg 
(e.g. \textit{Marley et al.} 2007); \textit{Spiegel and Burrows} 2012; \textit{Bonnefoy  et al.} 2013). Furthermore, it is possible that planets formed by core-accretion 
may not have stellar/solar abundances (\textit{Madhusudhan et al.} 2011b; \textit{\"{O}berg et al.} 2011).

A second approach is to rely exclusively on model atmosphere fits using a large
model grid that finely samples the free parameters that define
each model (\textit{Currie et al.}, 2011; \textit{Madhusudhan et al.}, 2011c). This method has 
the advantage of being completely independent of the evolution calculations and initial 
conditions and would allow a mapping of T$_{\rm eff}$ and $\log(g)$ as function of age for 
giant planets. In practice, however, atmosphere-only fitting can be difficult because the 
number of free parameters is fairly large (especially if basic composition is allowed to vary) 
and local minima are easily found when minimizing $\chi^2$.  Combining photometry across
many band passes and higher SNR data and higher spectral resolution are often
needed to find a reliable match.  Of course, giant planet atmospheres are
complex and model inadequacies also contribute to the poor and/or often
misleading fits.

\bigskip
\noindent
\textbf{5.5 The HR 8799 System}
\bigskip

The four planets orbiting the young star HR 8799, between 14 and 70 AU (0.3 to
1.7 arcseconds), are the best examples yet of what direct imaging can yield
(\textit{Marois et al.} 2008, 2010).  This system is providing new insights into the
formation and evolution of giant planets and currently serves as the best
system for testing planetary models and new instruments.  The HR 8799 system is
also an excellent early test of our ability to infer basic properties of
planets discovered by direct imaging. Unlike transits, when direct imaging is
possible, at least one flux measurement is immediately obtained. Thus,
planets found in this way immediately lend themselves to atmospheric
characterization.

The HR8799 bcd discovery included photometry covering $J$ through $L^\prime$
wavelengths and indicated that all three planets have red IR colors and,
therefore, traditional cloud-free atmosphere models are inappropriate (\textit{Marois
et al.}, 2008).  Marois et al. found that model atmospheres including
condensates suspended at pressures determined only by chemical equilibrium
systematically over predicted $T_{\rm eff}$, implying radii too small for the
well-determined $L_{\rm bol}$ and our current understanding of giant planet
formation. Marois et al.  also concluded that intermediate cloud models 
(constrained vertically) could simultaneously match the photometry and match
expectations from giant planet evolution tracks. The photometric quality and
wavelength coverage continues to improve (\textit{Hinz et al.}, 2010; \textit{Galicher et al.}, 
2011; \textit{Currie et al.}, 2011; \textit{Skemer et al.}, 2012) and the basic conclusions from
these data are that all four planets have $800 < T_{\rm eff} < 1200$K, have
cloud coverage that is comparable to the reddest L dwarfs (or even thicker,
\textit{Madhusudhan et al.}, 2011c), and that atmosphere models can easily overestimate
effective temperature unless additional, secondary, model parameters are
allowed to vary (see \textit{Marley et al.}, 2012 for a summary).  Most studies agree
that CO/CH$_4$ non-equilibrium chemistry is necessary to fit the photometry.
Observations at 3.3~$\mu$m, overlapping a broad CH$_4$ band, are
particularly sensitive to non-equilibrium chemistry (\textit{Bowler et al.}, 2010; 
\textit{Hinz et al.}, 2010; \textit{Skemer et al.}, 2012). 

In addition to photometry, the HR8799 planets have been studied
spectroscopically.  Measurements have been made across two narrow bands, from
3.9 -- 4.1 $\mu$m (HR8799c: \textit{Janson et al.}, 2010) and 2.12 -- 2.23 $\mu$m
(HR8799b: \textit{Bowler et al.}, 2010), as well as across the full $J$, $H$ and $K$
bands (HR8799b: \textit{Barman et al.}, 2011; HR8799bcd: \textit{Oppenheimer et al.} 2013;
HR8799c: \textit{Konopacky et al.}, 2013).  The $H$-band spectrum of HR8799b has a shape
indicative of low surface gravity, consistent with its low mass and youth.
Low-resolution spectra for HR8799b and c show only weak or no evidence for
CH$_4$ absorption; best explained by non-equilibrium chemistry, as already
suggested by photometry.

Unless initially constrained by evolution model predictions, model atmosphere
fits to the low-resolution spectroscopic data (e.g. for HR8799b) can easily
overestimate $T_{\rm eff}$.  This situation is identical to the one encountered
when fitting the photometry.  The challenge is to produce a cool model
atmosphere ($T_{\rm eff} <$ 1000K) that simultaneously has red near-IR colors,
has weak CH$_4$ and strong H$_2$O absorption matching the observed spectra, and
the right overall $H$ and $K$ band shapes.  When minimizing $\chi^2$, local
minima are often found because more than one physical property can redden the
spectrum, in addition to clouds.  As gravity decreases so does collision
induced H$_2$ opacity (which is strongest in the $K$-band), allowing more flux
to escape at longer wavelengths.  Also, uniform increases in metals
preferentially increases the opacities at shorter wavelengths.  So far, the
best approach for dealing with potential degeneracies has been to combine
photometry covering as much of the SED as possible (influenced mostly by
$T_{eff}$) with as much spectral information as possible (influenced mostly by
gravity and non-equilibrium chemistry).  Additional constraints on radius from
evolution models are also useful to keep the overall parameter space within
acceptable/physical boundaries.  As atmosphere models and observations improve,
some constraints will hopefully be unnecessary.

HR8799c is nearly a magnitude brighter than HR8799b, allowing moderate spectral
resolution observations ($R \sim 4000$) and has revealed hundreds of molecular lines from
H$_2$O and CO (\textit{Konopacky et al.}, 2013).  These data provided additional evidence
for low surface gravity and non-equilibrium chemistry. \textit{Konopacky et al.} 
detected no CH$_4$ lines which strongly supports the non-equilibrium CO/CH$_4$
hinted at by the photometry and low-resolution spectra.  More importantly,
access to individual lines allowed the C and O abundances to be estimated, with
the best-fitting C/O being larger than the host star, hinting at formation by
core-accretion (\textit{Konopacky et al.}, 2013; \textit{Madhusudhan et al.} 2011b; \textit{\"{O}berg et al.}, 2011).

\bigskip
\centerline{\textbf{6. ATMOSPHERES OF HOT NEPTUNES}} 
\bigskip

Exo-Neptunes are  loosely defined here as planets with masses and radii similar to those of the ice giants in the solar system ($M_p$ between 10 and 30 $M_\oplus$; $R_p$ between 1 and 5 $R_\oplus$). Masses and radii have been measured for six transiting exo-Neptunes to date. Since all these objects have  equilibrium temperatures of $\sim$ 600 - 1200 K, compared to $\sim$70 K for Neptune, they are referred to as `hot Neptunes'. While the interiors of ice giant planets are expected to be substantially enriched in ices and heavy elements, their atmospheres are generally expected to be H$_2$-rich. Their expected low mean-molecular weight together with their high temperatures make hot Neptunes conducive to atmospheric observations via transit spectroscopy just like hot Jupiters. On the other hand, their temperatures  are significantly lower than those of hot Jupiters ($\sim 1300 - 3300$ K), making them particularly conducive to study non-equilibrium processes in their atmospheres (see section 3.3). 

Atmospheric observations over multiple spectral bands have been reported for only one hot Neptune, GJ 436b, to date. \textit{Stevenson et al.} (2010) reported planet-star flux contrasts of the day-side atmosphere of GJ~436b in six channels of {\it Spitzer} broadband photometry, explaining which required a substantial depletion of CH$_4$ and overabundance of CO and CO$_2$ compared to equilibrium predictions assuming solar abundances (\textit{Stevenson et al.}, 2010;  \textit{Madhusudhan and Seager}, 2011; also see \textit{Spiegel et al.}, 2010). \textit{Knutson et al.} (2011) reported multiple photometric observations of GJ~436b in transit in each of the 3.6 \& 4.5 $\mu$m {\it Spitzer} channels, which indicated variability in the transit depths between the different epochs. For the same set of observations as \textit{Stevenson et al.} (2010) and \textit{Knutson et al.} (2011), \textit{Beaulieu et al.} (2010) used a subset of the data and/or adopted higher uncertainties, to suggest the possibility of a methane-rich atmosphere consistent with equilibrium predictions. On the other hand, \textit{Shabram et al.} (2011) reported models fits to the transit data and found their models to be inconsistent with those used by \textit{Beaulieu et al.} (2010). 

The inferences of low CH$_4$ and high CO in GJ~436b suggest extreme departures from equilibrium chemistry. As discussed in section~3.3, in low temperature atmospheres CH$_4$ and H$_2$O are expected to be abundant whereas CO is expected to be negligible. \textit{Madhusudhan and Seager} (2011) suggested that the inferred CO enhancement in GJ~436b can be explained by a combination of super-solar metallicity ($\geq 10 \times$ solar) and non-equilibrium chemistry, via vertical eddy mixing of CO from the hotter lower regions of the atmosphere. The required eddy mixing coefficient ($K_{zz}$) of $\sim 10^7$ cm$^2$/s  (\textit{Madhusudhan and Seager} 2011) is also consistent with that observed in 3D General Circulation Models of GJ~436b (\textit{Lewis et al.} 2010). However, even though CH$_4$ can be photochemically depleted by factors of a few in the upper atmospheres of irradiated giant planets (e.g. \textit{Moses et al.}, 2011), the drastic depletion required in the observable lower atmosphere (at the 100 mbar pressure level) of GJ~436b has no explanation to date. Furthermore, \textit{Shabram et al.} (2011) suggested that at the low temperatures of GJ~436b, other higher-order hydrocarbons such as HCN and C$_2$H$_2$ can also be abundant in the atmosphere, and contribute spectral features overlapping with those of CH$_4$. Recently, \textit{Moses et al.}~(2013) suggested an alternate explanation that a CH$_4$-poor and CO-rich composition in GJ~436b could potentially be explained by an extremely high atmospheric metallicity ($\sim1000 \times$ solar). Future observations with better spectral resolution and wider wavelength coverage would be needed to refine the molecular abundances and constrain between  the different hypotheses. 

\begin{figure*}[ht]
\centering
\includegraphics[width= \textwidth]{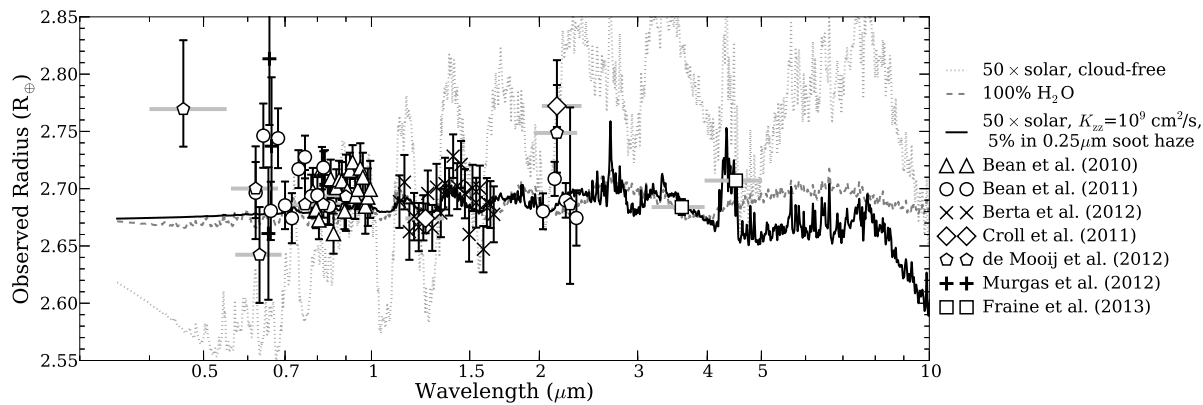}
\caption{Model transmission spectra and observations of the super-Earth GJ1214b.  In dotted light gray is a transmission model with a metallicity 50$\times$ solar.  In dashed dark gray is 100\% steam atmosphere.  In black is a 50$\times$ solar model that also includes a hydrocarbon haze layer from \emph{Morley et al.} (2013).  The cloud-free 50$\times$ model is clearly ruled out by the data, but with current data the steam and hazy model give very similar quality fits.  Future high S/N data in the near infrared, or especially the mid-infrared, with help to distiguish between models.}
\label{fig:gj1214b}
\end{figure*}

\bigskip
\centerline{\textbf{7. ATMOSPHERES OF SUPER-EARTHS}} 
\bigskip

``Super-Earths" are nominally defined as planets with masses between 1 and 10 \me. Super-Earths have no analogues in the solar system, their masses being intermediate between those of terrestrial planets and ice giants. As such, it is presently unknown if super-Earths are mini-Neptunes with H/He-rich atmospheres, or are scaled up terrestrial planets with atmospheres dominated by heavy molecules (H$_2$O, CO$_2$, etc.). It is also unclear if the heavier elements beneath the atmospheres are predominantly rock/iron, or if they possess a large mass fraction of volatile ices, as has long been suggested for Uranus and Neptune. Their being the lowest mass planets whose atmospheres can potentially be observed with existing and upcoming instruments, understanding super-Earths is a major frontier in exoplanetary science today. In the present section, we focus on the atmospheres of volatile-rich super-Earths, such as GJ~1214b.

\bigskip
\noindent
\textbf{7.1 Theory}
\bigskip

Planetary atmospheres can be classified as either ``primary'' or ``secondary.''  Primary atmospheres are those that are accreted directly from the nebula, and are therefore dominated by hydrogen and helium gas.  Secondary atmospheres are those that are made up of volatiles outgassed from the planet's interior.  Within the solar system, Jupiter, Saturn, Uranus, and Neptune have primary atmospheres, while the atmospheres of the smaller bodies (rocky planets and moons) are secondary atmospheres.  It is possible that Uranus and Neptune, which are only 10-20\% H-He by mass, could have some minor secondary component.

There is a strong expectation that planets along the continuum from 1-15 \me\ will posses diverse atmospheres that could either be predominantly primary or predominantly secondary or mixed.  The accretion of primary atmospheres has long been modeled within the framework of the core-accretion theory for planet formation (see the chapter by Helled et al.).  The outgassing of secondary atmospheres has long been modeled in the solar system, but relatively little work has occurred for exoplanets (\emph{Elkins-Tanton and Seager} 2008, \emph{Rogers and Seager} 2010).  One particular finding of note is that hydrogen atmospheres up to a few percent of the planet's mass could be outgassed.  Unlike primary atmospheres, these would be helium free. A primary future goal of characterizing the atmospheres of planets in the super-Earth mass range is to understand the extremely complex problem of the relative importance of primary and secondary atmospheric origin for planets in this mass range. 

The tools to observe super-Earth atmospheres will be the same used for gas giants:  initially transmission and emission spectroscopy, and eventually direct imaging. A number of recent studies have reported atmospheric models of super-Earths to aid in the interpretation of observed spectra. These studies investigated theoretical spectra and retrieval methods (\emph{Miller-Ricci et al.} 2009; \emph{Miller-Ricci and Fortney}, 2010; \emph{Benneke and Seager} 2012; \emph{Howe and Burrows} 2012), atmospheric chemistry (e.g. \emph{Kempton} et al. 2011), clouds and hazes (\emph{Howe and Burrows} 2012; \emph{Morley et al} 2013), for super-Earth atmospheres.

Spectroscopic identification of abundant gases in the planetary atmosphere would be the clearest and most direct path to assessing bulk atmospheric composition. A more indirect route would come from measurements of the scale height of a transiting super-Earth atmosphere in any one molecular band. \emph{Miller-Ricci et al.} (2009) pointed out that a measurement of the atmospheric scale height from a transmission spectrum would be a direct probe of the atmospheric mean molecular weight (MMW) and hence the bulk composition of a super-Earth. Atmospheres with larger scale heights yield more dramatic variation in the transit radius as a function of wavelength. In principle, H-rich atmospheres could be easily separated from those dominated by heavier molecules like steam or carbon dioxide.  However, it is important to keep in mind that clouds, which are generally gray absorbers or scatterers, could also obscure transmission spectrum features (\emph{Fortney} 2005; \emph{Howe and Burrows} 2012; \emph{Morley et al} 2013).  However, transmission spectrum observations with a high enough signal-to-noise should readily be able to recover a wealth of atmospheric information (\emph{Benneke \& Seager} 2012).

\bigskip
\noindent
\textbf{7.2 GJ 1214b as Prototype}
\bigskip

The super-Earth GJ 1214b has been the target of many observational campaigns to better understand the transmission spectrum from the blue to the mid infrared.  From a mass and radius measurement alone, the composition is degenerate (\emph{Rogers \& Seager} 2010, \emph{Nettelmann et al.} 2011, \emph{Valencia et al.} 2013).  If the outer layer of the planet, including the visible atmosphere, can be probed we will have a much better understanding of the planet's bulk composition (\emph{Miller-Ricci \& Fortney}, 2010).  A cloud-free atmosphere of solar composition, with a relatively large scale height, was quickly ruled out by \emph{Bean  et al.} (2010). Subsequent data from {\it Spitzer} (\emph{D\'esert et al.} 2011c) and {\it HST} WFC3 (\emph{Berta et al.} 2012a), along with ground-based data, concurred with this view. With these data, the transmission spectrum was essentially flat (Fig.~\ref{fig:gj1214b}), and was consistent with either a high-MMW atmosphere or an atmosphere blanketed by opaque clouds. Most recently, very high S/N data with {\it HST} WFC3 still show a flat spectrum ruling out a cloud-free high MMW atmosphere (Kriedberg et al. 2014). 

As has been suggested for GJ 436b, it may be that the atmosphere is \emph{strongly} super-solar in abundances, perhaps hundreds of times solar, or even higher.  \emph{Fortney et al.} (2013) suggest from population synthesis models that metal rich envelopes for low-mass $\sim$5-10 \me\ planets may commonly reach values of $Z$ from 0.6-0.9.  Such atmospheres would naturally appear more featureless in transmission spectra due to a smaller scale height and would have abundant material to form clouds and hazes.

With only one example yet probed, we are not yet sure if the difficulty in obtaining the transmission spectrum of GJ 1214b will be normal, or will be the exception. Photometric detections of transits, and an eclipse, in the {\it Spitzer} bandpasses have also been reported for another super-Earth 55 Cancri e (\emph{Demory et al.}, 2012; \emph{Gillon et al.}, 2012), but the planet-star radius ratio in this case is significantly smaller compared to GJ~1214b making spectral observations challenging. Recently another transiting planet, HD 97658b, of a similar mass and radius to GJ~1214b, was discovered around a bright K-star (\emph{Dragomir et al.}, 2013), presenting a promising candidate for spectroscopic follow up and comparison to GJ~1214b. The ubiquity of 2-3 \re\ planets on relatively short periods orbits around M stars and Sun-like stars suggests we will eventually have a large comparison sample of these volatile-rich low-mass planets.

\bigskip
\centerline{\textbf{8. FUTURE OUTLOOK}}
\bigskip

Observations with existing and upcoming facilities promise a bright outlook for the characterization of exoplanetary atmospheres. Currently, low resolution data from space-borne ({\it HST} and {\it Spitzer}) and ground-based instruments are already leading to nominal inferences of various atmospheric properties of exoplanets. Future progress in atmospheric observations and theory can advance exoplanetary characterization in three distinct directions, depicted in Fig.~1. Firstly, robust inferences of chemical and thermal properties will enhance our understanding of the various physico-chemical processes in giant exoplanetary atmospheres discussed in this chapter. Secondly, better constraints on the elemental abundance ratios in such atmospheres could begin to inform us about their formation environments. Finally, high precision observations could help constrain the atmospheric compositions of super-Earths, which, in turn, could constrain their interior compositions. Several new observational facilities on the horizon will aid in these various aspects of exoplanetary characterization. 

\bigskip
\noindent
\textbf{8.1 Next-generation Atmospheric Observations of Transiting Giant Planets}

\bigskip

Over the next decade the development of large-aperture telescopes such as the  European Extremely Large Telescope  (E-ELT), the Giant Magellan Telescope  (GMT), and the Thirty Meter Telescope  (TMT) will open up new opportunities for studies of both transiting and non-transiting planets.  The increased aperture of these telescopes will provide an immediate advantage for transmission spectroscopy and phase variation studies using visible and near-infrared echelle spectroscopy, which are currently photon-noise-limited even for the brightest stars (e.g., \emph{Redfield et al.} 2008; \emph{Jensen et al.} 2011, 2012; \emph{Snellen et al.} 2010b; \emph{Brogi et al.} 2013; \emph{de Kok et al.} 2013).  Such facilities could also make it feasible to characterize the atmospheres of terrestrial-size exoplanets (\emph{Snellen et al.} 2013; \emph{Rodler and L\'opez-Morales} 2014). It is less clear whether or not such large telescopes will provide good opportunities for broadband photometry and low-resolution spectroscopy of transits and secondary eclipses, as this technique requires a large field of view containing multiple comparison stars in order to correct for time-varying telluric absorption and instrumental effects (e.g., \emph{Croll et al.} 2010, 2011).  Most broadband studies using this technique achieve signal-to-noise ratios that are a factor of 2-3 above the photon noise limit, implying that increased aperture may not lead to a corresponding reduction in noise until the limiting systematics are better understood. 

The James Webb Space Telescope ({\it JWST}; \textit{Garder et al.},~2006) currently offers the best future prospects for detailed studies of exoplanetary  atmospheres. Given its large aperture and spectral coverage, {\it JWST} will be capable of obtaining very high S/N and high resolution spectra of transiting giant planets, thereby rigorously constraining their chemical and thermal properties. In particular, the spectroscopic  capabilities of the NIRSpec (0.6-5.0 $\mu$m) and MIRI (5-28 $\mu$m) instruments aboard {\it JWST} will provide a wide wavelength coverage including spectral features of all the prominent molecules in gas giant atmospheres (see section 3.3). On the other hand, the possibility of dedicated space missions like EChO for atmospheric characterization of exoplanets (Barstow et al. 2013; Tinetti et al. 2013) will significantly increase the sample of giant exoplanets with high quality spectral data. 

\bigskip
\noindent
\textbf{8.2 The Small Star Opportunity} 
\bigskip

If we wish to study the properties of smaller and more earth-like planets, it is crucial to find these objects orbiting small, nearby stars.  The reason for this can be easily understood in light of Equations \ref{transm_eq} and \ref{sec_ecl_eq}.  As the size and temperature of the planet decrease the corresponding transmission spectrum amplitude and secondary eclipse depth will also drop rapidly.  If we evaluate Equation \ref{transm_eq} for the earth-sun system, we find that the earth's atmosphere has a scale height of approximately 8.5 km at a temperature of 290 K and assuming a mean molecular weight of $4.81\times10^{-26}$ kg.  The corresponding depth of the absorption features in earth's transmission spectrum is then $2\times10^{-6}$, or a factor of $\sim$1000 smaller than the signal from the transiting hot Jupiter HD 189733b.  If we carry out the same calculation for a secondary eclipse observation using Equation \ref{sec_ecl_eq}, we find a predicted depth of approximately $4\times10^{-6}$, also a factor of $\sim$1000 smaller than for HD 189733b.  Rather than building a space telescope with a collecting area a million times larger than existing facilities, we can mitigate the challenges inherent in measuring such tiny signals by focusing on observations of small planets transiting small stars. 

If we instead place our transiting earth in front of a late-type M5 dwarf with a radius of 0.27 R$_{\sun}$ and an effective temperature of 3400 K, we gain a factor of 10 for the transmission spectrum and a factor of 25 for the secondary eclipse depth.  These kinds of signals, although more favorable, may still be beyond the reach of the \emph{JWST} (\emph{Greene et al.} 2007; \emph{Kaltenegger and Traub} 2009; \emph{Seager et al.} 2009).  If we are instead willing to consider super-earth-sized planets, with lower surface gravities and larger scale heights than Earth, it may be possible to measure transmission spectra for these more favorable targets.  Similarly, if we extend our secondary eclipse observations to super-Earths, which have larger radii and hence deeper secondary eclipses, such observations could be achievable with \emph{JWST}.

There are currently three small ground-based transiting planet surveys that are focused exclusively on searching the closest, brightest M dwarfs (\emph{Nutzman and Charbonneau} 2008; \emph{Berta et al.} 2012b; \emph{K\'ovacs et al.} 2013; \emph{Sozzetti et al.} 2013).  To date only one survey, MEarth, has detected a new transiting super-Earth (\emph{Charbonneau et al.} 2009), but as these surveys mature it is likely that additional systems will be discovered.  Another promising avenue is to search for transits of known low-mass planets detected by radial velocity surveys; two transiting super-earths to date have been detected this way (\emph{Winn et al.} 2011; \emph{Demory et al.} 2011b; \emph{Dragomir et al.} 2013).  However, neither of these surveys are expected to yield large numbers of low-mass transiting planets orbiting M stars, due to a combination of limited sample sizes and sensitivity limits that are too high to detect very small planets.  If we wish to detect large numbers of transiting planets it will require a new space mission that will survey all the brightest stars in the sky with a sensitivity high enough to detect transiting super-Earths. The Transiting Exoplanet Survey Satellite (TESS) mission was recently approved  by NASA for launch in 2017, and currently represents our best hope for building up a large sample of small transiting planets suitable for characterization with \emph{JWST}.  In addition, the CHaracterizing ExOPlanet Satellite (CHEOPS) was recently selected by ESA for launch in 2017, which will search for transits of small planets around bright stars that are already known to host planets via radial velocity searches. 

\bigskip
\noindent
\textbf{8.3 Expectations for New Direct Imaging Instruments} 
\bigskip

New specialized direct-imaging instruments are, or soon will be, available at
ground-based telescopes behind newly designed adaptive optics systems.  These
 include the Spectro-Polarimetric High-contrast Exoplanet REsearch (SPHERE)
instrument for the European Southern Observatory's Very Large Telescope
(\textit{Beuzit et al.}, 2006), Project 1640 (P1640) for Palomar Observatory
(\textit{Oppenheimer et al.}, 2013), the Gemini Planet Imager (GPI) for the Gemini
South telescope (\textit{Macintosh et al.}, 2008), , the PISCES and LMIRCam facilities for the Large Binocular Telescope (LBT) (e.g. \textit{Skemer et al.}, 2012), and the HiCIAO instrument for the Subaru telescope for the SEEDS project (e.g. \textit{Tamura} 2009; \textit{Kuzuhara et al.}, 2013). These instruments are optimized to reduce 
scattered light within an annulus centered on the star, sacrificing good
contrast at very wide separations in favor of a high-contrast field of view
optimized for sub-Jupiter-mass planets in the 5 to 50 AU range. Any candidate
within this region will have near-IR fluxes and low-resolution ($R \sim 45$)
spectra measured (e.g. the $H$ band).  Surveys with these instruments will
determine the frequency of Jovian-mass planets for a range of stellar types and
ages, in wider orbits than are available to RV and transit methods.  For a subset
of the discoveries, low-resolution spectra, at high-SNR, will be obtained across
most of $YJHK$ band (depending on the instrument).   Additional mid-IR
observations from the ground will be possible, e.g, with the Magellan
AO system providing valuable flux measurements closer to the SED peak
for the coldest, lowest mass, direct imaging discoveries (\textit{Close et al.}, 2012). 

The prospects for atmosphere characterization are excellent with access to so 
large an amount of spectral data.  Even at low resolution, broad molecular
features from H$_2$O and CH$_4$ should be measurable.  If clouds are present,
the spectra will be noticeably smoother and redder, as discussed in section~5. 
Having access to $Y$ and $J$ bands will greatly aid in narrowing in on the preferred
range of cloud particle sizes and thicknesses. $H$ and $K$-band spectra will
provide independent estimates of surface gravity.  As learned from
HR8799, model atmospheres continue to prove problematic for inferring \teff\
and surface gravity.  However, having spectra for many planets with different
ages, masses, and luminosities will quickly lead to improvements in theory.

Giant ground-based telescopes ($\sim$ 30m-class) and new instrumentation in
space promise to greatly increase the number of directly imaged planetary
systems, with greater contrasts and down to lower masses and/or to older ages.
JWST, with IR and mid-IR capabilities (NIRCAM, MIRI and TFI), will provide much
needed longer wavelength capability for direct spectroscopy of
sub-Jovian planets. 

\noindent
\textbf{Acknowledgments.} NM acknowledges support from Yale University through the YCAA fellowship. TB acknowledges support from NASA awards NNX10AH31G, NNH10AO07I and NSF award AST-1405505. JJF acknowledges support from NASA awards NNX09AC22G, NNX12AI43A and NSF award AST-1312545.

\bigskip
\noindent
\textbf{REFERENCES}

\refs Agol E., et al. (2005) \emph{Astrophys. J., 359}, 567-579.
\refs Agol E., et al. (2010) \emph{Astrophys. J., 721}, 1861-1877.
\refs Alonso R., et al. (2009a) \emph{Astron. Astrophys., 501}, L23-L26.
\refs Alonso R., et al. (2009b) \emph{Astron. Astrophys., 506}, 353-358.
\refs Anderson D. R. et al. (2010), \emph{Astron. Astrophys., 513}, L3, 5pp.
\refs {Anderson} D. R. et al. (2013) \emph{Mon. Not. R. Astron. Soc., 430}, 3422-3431.
\refs Asplund M., et al. (2009) \textit{Annual Rev. Astron. Astrophys., 47}, 481-522.
\refs Astudillo-Defru N. and Rojo P. (2013) \textit{Astron. Astrophys., 557}, A56, 8pp 
\refs {Bailey} V., et al. (2014) \emph{Astrophys. J. Lett., 780}, L4, 6pp. 
\refs {Baines} E.~K., et al. (2012) \emph{Astrophys. J.}, \emph{761}, 57.
\refs {Baraffe} I., et al.. (2003) \emph{Astron. Astrophys.}, \emph{402}, 701-712.
\refs {Barclay} T., et al. (2012) \emph{Astrophys. J.}, \emph{761}, 53.
\refs Barman T. S., Hauschildt P. H. and Allard, F. (2005) \textit{Astrophys. J., 632}, 1132-1139. 
\refs Barman T. S. (2007) \textit{Astrophys. J., 661}, L191-L194.  
\refs Barman T. S. (2008) \textit{Astrophys. J., 676}, L61-L64. 
\refs {Barman} T. S., et al. (2011) \emph{Astrophys. J.}, \emph{733}, 65. 
\refs {Barstow} J. K., et al. (2013) \emph{Mon. Not. R. Astron. Soc., 430}, 1188-1207.
\refs {Bean} J.~L., {Miller-Ricci Kempton} E., and {Homeier} D. (2010) \emph{Nature},
  \emph{468}, 669--672.
\refs Bean J. L., et al. (2011) \emph{Astrophys. J., 743}, 92-104.
\refs Bean J. L., et al. (2013) \emph{Astrophys. J., 771}, 108-119.
\refs Beaulieu J. -P., et al. (2008) \emph{Astrophys. J., 677}, 1343-1347. 
\refs Beaulieu J. -P., et al. (2011) \emph{Astrophys. J., 731}, 16-27. 
\refs {Becklin} E.~E. and {Zuckerman}, B. (1988) \emph{Nature}, \emph{336}, 656.
\refs {Benneke} B. and {Seager} S. (2012) \emph{Astrophys. J.}, \emph{753}, 100.
\refs Berdyugina S. V., et al. (2008) \emph{Astrophys. J., 673}, L83-L86. 
\refs Berta Z. K. et al., (2012a) \emph{Astrophys. J., 747}, 35-52.
\refs Berta Z. K. et al. (2012b) \emph{Astron. J., 144}, 145-165.
\refs Beuzit J. -L. et al. (2006) \emph{The Messenger, 125}, 29-34. 
\refs Birkby J. L. et al. (2013) \emph{Mon. Not. R. Astron. Soc. Lett., 436}, L35-L39. 
\refs Bonnefoy M. et al. (2013) \emph{Astron. Astrophys., 555}, A107, 19pp.
\refs Borucki W. J. et al. (2009) \emph{Science, 325}, 709-709.
\refs Bouchy F. et al. (2005) \emph{Astron. Astrophys., 444}, L15-L19.
\refs {Bowler} B.~P., et al.. (2010) \emph{Astrophys. J.}, \emph{723}, 850-868.
\refs Brogi M. et al. (2012) \emph{Nature, 486}, 502-504. 
\refs Brogi M. et al. (2013) \emph{Astrophys. J., 767}, 27-37.
\refs Brown T. M. (2001) \textit{Astrophys. J., 553}, 1006-1026. 
\refs Burrows A. and Sharp C. M. (1999) \textit{Astrophys. J., 512}, 843-863. 
\refs Burrows A., et al. (2007) \textit{Astrophys. J., 668}, L171-L174. 
\refs Burrows A. et al. (2008a) \emph{Astrophys. J., 682}, 1277-1282.
\refs Burrows A., Budaj J. and Hubeny I. (2008b) \textit{Astrophys. J., 678}, 1436-1457. 
\refs Burke C. J. et al. (2010) \emph{Astrophys. J., 719}, 1796-1806.
\refs Campo C., et al. (2011) \emph{Astrophys. J., 727}, 125-136. 
\refs {Carson} J., et al. (2013) \emph{Astrophys. J.}, \emph{763}, L32--37.
\refs Charbonneau D., et al. (2002) \emph{Astrophys. J., 568}, 377-384.
\refs Charbonneau D., et al. (2005) \emph{Astrophys. J., 626}, 523-529.
\refs Charbonneau D., et al. (2008) \emph{Astrophys. J., 686}, 1341-1348.
\refs Charbonneau D., et al. (2009) \emph{Nature, 462}, 891-894.
\refs Chauvin G., et al. (2005) \emph{Astron. Astrophys., 438}, L25-L28. 
\refs Cho J., et al. (2008) \emph{Astrophys. J., 675}, 817-845.
\refs Christiansen J. L., et al. (2010) \emph{Astrophys. J., 710}, 97-104.
\refs Close L. M., et al. (2012) \emph{Proc. SPIE, 8447}, 84470X, 16pp.
\refs Cooper C. S. and Showman A. P. (2006) \textit{Astrophys. J., 649}, 1048-1063. 
\refs Coughlin J. L. and L\'opez-Morales M. (2012) \emph{Astron. J., 143}, 39-61.
\refs Cowan N., et al. (2007) \emph{Mon. Not. R. Astron. Soc., 379}, 641-646.
\refs Cowan N., and Agol, E. (2008) \emph{Astrophys. J., 678}, L129-L132.
\refs Cowan N., and Agol, E. (2011a) \emph{Astrophys. J., 726}, 82-94.
\refs Cowan N., and Agol, E. (2011b) \emph{Astrophys. J., 729}, 54-65.
\refs Cowan N., et al. (2012) \emph{Astrophys. J., 747}, 82-99.
\refs Croll B., et al. (2010) \emph{Astrophys. J., 717}, 1084-1091.
\refs Croll B., et al. (2011) \emph{Astronom. J., 141}, 30-42.
\refs Crossfield I. J. M., et al. (2012a) \emph{Astrophys. J., 752}, 81-94.
\refs Crossfield I. J. M., Hansen B., and Barman, T. (2012b) \emph{Astrophys. J., 746}, 46, 17 pp.
\refs Crossfield I. J. M. (2012c) \emph{Astrophys. J., 760}, 140-156.
\refs Crouzet N. et al. (2012) \emph{Astrophys. J., 761}, 7-20.
\refs Cumming A., et al. (2008) \emph{Publ. Astron. Soc. Pacific, 120}, 531-554. 
\refs {Currie} T., et al. (2011) \emph{Astrophys. J.}, \emph{729}, 128-147. 
\refs {Cushing} M.~C., et al. (2011) \emph{Astrophys. J.}, \emph{743}, 50.
\refs de Kok R., Stam D., Karalidi T. (2011) \emph{Astrophys. J., 741}, 59, 6pp.
\refs de Kok R., et al. (2013) \emph{Astron. Astrophys., 554}, A82-A91.
\refs de Mooij E. et al. (2012) \emph{Astron. Astrophys., 538}, 46-58.
\refs de Wit J. et al. (2012) \emph{Astron. Astrophys., 548}, A129-A148.
\refs Deming, D. et al. (2005) \emph{Nature, 434}, 738-740.
\refs Deming D. et al. (2013) \emph{Astrophys. J., 774,} 95-111.
\refs Demory B. et al. (2011a) \emph{Astrophys. J., 735}, L12-L18.
\refs Demory B. et al. (2011b) \emph{Astron. Astrophys., 533}, A114-121.
\refs Demory B. et al. (2012) \emph{Astrophys. J. Lett., 751}, L28, 7pp.
\refs Demory B. et al. (2013) \emph{Astrophys. J. Lett., 776}, L25, 7pp.
\refs D\'esert J.-M., et al. (2009) \emph{Astrophys. J., 699}, 478-485.
\refs D\'esert J.-M., et al. (2011a) \emph{Astrophys. J., 197}, 14-27.
\refs D\'esert J.-M., et al. (2011b) \emph{Astrophys. J. Supp., 197}, 11-21.
\refs D\'esert J.-M., et al. (2011c) \emph{Astrophys. J., 731}, L40.
\refs Dobbs-Dixon I., Cumming A., Lin D. N. C. (2010) \emph{Astrophys. J., 710}, 1395--1407
\refs Dragomir D. et al. (2013) \emph{Astrophys. J. Lett., 772}, L2, 6pp.
\refs {Elkins-Tanton} L.~T. and {Seager} S. (2008) \emph{Astrophys. J.}, \emph{685},
  1237--1246.
\refs Esteves L. J., De Mooij E. J. W. and Jayawardhana R. (2013) \emph{Astrophys. J., 772}, 51-64.
\refs Evans T. M., et al. (2013) \emph{Astrophys. J., 772}, L16-L20.
\refs {Fabrycky}, D.~C. and {Murray-Clay}, R.~A. (2010) \emph{Astrophys. J.}, \emph{710}, 1408-1421.
\refs Fegley B. and Lodders K. (1994) \textit{Icarus, 110}, 117-154. 
\refs {F\"{o}hring} D. et al. (2013) \emph{Mon. Not. R. Astron. Soc., 435}, 2268-2273.
\refs Fortney J. (2005) \emph{Mon. Not. R. Astron. Soc., 364}, 649-653.
\refs {Fortney} J.~J., et al. (2006) \emph{Astrophys. J.}, \emph{642}, 495--504.
\refs {Fortney}, J.~J., et al. (2007) \emph{Astrophys. J.}, \emph{659}, 1661--1672.
\refs {Fortney}, J.~J., et al. (2008a) \emph{Astrophys. J.}, \emph{678}, 1419--1435.
\refs {Fortney}, J.~J., et al. (2008b) \emph{Astrophys. J.}, \emph{683}, 1104--1116. 
\refs {Fortney} J.~J., et al. (2013) \emph{Astrophys. J.}, \emph{775}, 80--92. 
\refs Fossati L., et al., (2010) \emph{Astrophys. J., 714}, L222-L227.
\refs {Fraine} J.~D., et al. (2013) \emph{Astrophys. J.}, \emph{765}, 127--139. 
\refs Freedman R. S., Marley M. S., and Lodders, K. (2008) \textit{Astrophys. J. Supp., 174},504-513.
\refs {Galicher} R. et al. (2011) \emph{Astrophys. J.}, \emph{739}, L41.
\refs {Gardner} J. P. et al. (2006) \emph{Space Sci. Rev.}, \emph{123}, 485-606. 
\refs Gibson N. P. et al. (2011)  \emph{Mon. Not. R. Astron. Soc., 411}, 2199-2213.
\refs Gibson N. P. et al. (2012) \emph{Mon. Not. R. Astron. Soc., 422}, 753-760.
\refs Gillon M. et al. (2012) \emph{Astron. Astrophys., 539}, A28-A34.
\refs Gillon M. et al. (2012) \emph{Astron. Astrophys., 542}, A4-A18.
\refs Goody R. M. and Yung Y. (1989) in \textit{Atmospheric radiation : theoretical basis}, ed. R. M. Goody and Y. Yung, New York, NY: Oxford University Press, 1989.
\refs Greene T. et al. (2007) \emph{Proc. SPIE, 6693}, 1-10.
\refs Grillmair C. J. et al. (2008) \emph{Nature, 456}, 767-769.
\refs Guillot T. (2010) \emph{Astron. Astrophys., 520}, A27, 13pp.
\refs Harrington J. et al. (2006) \emph{Science, 314}, 623-626.
\refs Hansen B. (2008) \textit{Astrophys. J. Supp., 179}, 484-508.
\refs Helling Ch., et al. (2008) \textit{Astron. Astrophys., 485}, 547-560. 
\refs Helling Ch. and Lucas, W. (2009) \textit{Mon. Not. R. Astron. Soc., 398}, 985-994.
\refs Heng K. et al. (2011) \textit{Mon. Not. R. Astron. Soc., 418}, 2669-2696.
\refs Heng K. et al. (2012) \textit{Mon. Not. R. Astron. Soc., 420}, 20-36.
\refs {Hinkley} S., et al. (2013) \emph{Astrophys. J.}, \emph{779}, 153--167.
\refs Hinz P. M., et al. (2010) \textit{Astrophys. J., 716}, 417-426.
\refs Holman M. J. and Murray, N. W. (2005)  \emph{Science, 307}, 1288-1291.
\refs Howe A.~R. and Burrows A.~S. (2012) \emph{Astrophys. J., 756}, 176-189
\refs Hubbard W. B. et al. (2001) \emph{Astrophys. J. 560}, 413-419.
\refs {Hubeny} I., {Burrows} A., and {Sudarsky} D. (2003) \emph{Astrophys. J.}, \emph{594},
  1011--1018.
\refs Huitson C. M. et al. (2012) \emph{Mon. Not. R. Astron. Soc., 422}, 2477-2488.
\refs Huitson C. M. et al. (2013) \emph{Mon. Not. R. Astron. Soc., 434}, 3252-3274.
\refs Janson M. et al. (2010) \emph{Astrophys. J., 710}, L35-L38. 
\refs Janson M. et al. (2013) \emph{Astrophys. J. Lett., 778}, L4-L9. 
\refs Jensen A. G. et al. (2011) \emph{Astrophys. J., 743}, 203-217.
\refs Jensen A. G. et al. (2012) \emph{Astrophys. J., 751}, 86-102.
\refs {Kalas} P., et al. (2008) \emph{Science}, \emph{322}, 1345.
\refs Kaltenegger L. and Traub, W. A. (2009) \emph{Astrophys. J., 698}, 519-527.
\refs Kempton E. M.-R. and Rauscher, E. (2012) \emph{Astrophys J., 751}, 117-129.
\refs Kipping D. M. and Spiegel, D. S. (2011) \emph{Mon. Not. R. Astron. Soc., 417}, L88-L92.
\refs Kipping D. M. and Bakos, G. (2011) \emph{Astrophys. J., 733}, 36-53.
\refs Knutson H. A. et al. (2007) \emph{Nature, 447}, 183-186.
\refs Knutson H. A. et al. (2008) \emph{Astrophys. J., 673}, 526-531.
\refs Knutson H. A. et al. (2009a) \emph{Astrophys. J., 690}, 822-836.
\refs Knutson H. A. et al. (2009b) \emph{Astrophys. J., 703}, 769-784.
\refs Knutson H. A. et al. (2010) \emph{Astrophys. J., 720}, 1569-1576.
\refs Knutson H. A. et al. (2011) \emph{Astrophys. J., 735}, 27-49.
\refs Knutson H. A. et al. (2012) \emph{Astrophys. J., 754}, 22-38.
\refs {Konopacky} Q.~M. et al. (2013) \emph{Science}, \emph{339}, 1398-1401. 
\refs Kopparapu R., Kasting, J. F., and Zahnle K. J.  (2012) \emph{Astrophys. J., 745}, 77-86.
\refs Kopparapu R. et al. (2013) \emph{Astrophys. J., 765}, 131-146.
\refs K\'ovacs et al. (2013) \emph{Mon. Not. R. Astron. Soc., 433}, 889-906. 
\refs {Kreidberg} L. et al. (2014) \emph{Nature}, \emph{505}, 7481, 69-72.
\refs {Kuzuhara} M., et al. (2013) \emph{Astrophys. J.}, \emph{774}, 11--28.
\refs {Lafreni{\`e}re} D., {Jayawardhana} R., and {van Kerkwijk} M.~H. (2008) \emph{Astrophys. J.,} \emph{689}, L153-L156. 
\refs {Lafreni{\`e}re} D., {Jayawardhana} R., and {van Kerkwijk} M.~H. (2010) \emph{Astrophys. J.,} \emph{719}, 497.
\refs {Lagrange} A.-M., et al. (2010) \emph{Science}, \emph{329}, 57.
\refs {Lagrange} A.-M., et al. (2009) \emph{Astron. Astrophys., 493}, L21-L25. 
\refs Lammer H. et al. (2003) \emph{Astrophys. J., 598}, L121-L124.
\refs Laughlin G. et al. (2009) \emph{Nature, 457}, 562-564.
\refs {Lecavelier des Etangs} A. et al., (2008) \emph{Astron. and Astrophys.}, \emph{481}, L83--L86.
\refs Lecavelier des Etangs A. et al. (2010) \emph{Astron. and Astrophys., 514}, A72-A82.
\refs Lee J.-M.,  Fletcher L. N., and Irwin, P. G. J. (2012) \emph{Mon. Not. R. Astron. Soc., 420}, 170-182.
\refs Lee J.-M.,  Heng K., and Irwin, P. G. J. (2013) \emph{Astrophys. J., 778}, 97-115.
\refs Lewis N. K., et al. (2010)  \textit{Astrophys. J., 720}, 344-356.
\refs Lewis N. K. et al. (2013) \emph{Astrophys. J., 766}, 95-117.
\refs Line M. R. et al. (2012) \emph{Astrophys. J., 749}, 93-102.
\refs Line M. R, Liang M-C, and Yung Y. L. (2010) \textit{Astrophys. J., 717}, 496-502. 
\refs Linsky, J. L. et al. (2010) \emph{Astrophys. J., 717}, 1291-1299.
\refs Lithwick Y., Xie J., and Wu Y. (2012)  \emph{Astrophys. J., 761}, 122-133.
\refs Lodders K. and Fegley B.  (2002) \textit{Icarus, 155}, 393-424. 
\refs Lodders K. (2004) \textit{Astrophys. J., 611}, 587-597.
\refs L\'opez-Morales M., et al. (2010) \textit{Astrophys. J., 716}, L36-L40.
\refs Machalek P. et al. (2008) \textit{Astrophys. J., 684}, 1427-1432.
\refs Macintosh B. et al. (2008) \textit{Proc. SPIE, 7015}, 701518, 13pp. 
\refs Madhusudhan N. and  Seager S. (2009) \textit{Astrophys. J., 707}, 24-39.
\refs Madhusudhan N. and Seager S. (2010) \textit{Astrophys. J., 725}, 261-274.
\refs Madhusudhan N. and Seager S. (2011) \textit{Astrophys. J., 729}, 41-54.
\refs Madhusudhan N., et al. (2011a) \textit{Nature, 469}, 64-67.
\refs Madhusudhan N., et al. (2011b) \textit{Astrophys. J., 743}, 191-202.
\refs {Madhusudhan} N., Burrows A., and Currie T. (2011c) \emph{Astrophys. J.}, \emph{737}, 34-48.
\refs Madhusudhan N. and Burrows, A. (2012) \emph{Astrophys. J., 747}, 25-41.
\refs {Madhusudhan} N. (2012) \emph{Astrophys. J.}, \emph{758}, 36.
\refs Majeau C. et al. (2012) \emph{Astrophys. J., 747}, L20-L25.
\refs Mancini L. et al. (2013), \emph{Mon. Not. R. Astron. Soc., 436}, 2-18
\refs Mandell A. et al. (2011) \emph{Astrophys. J., 728}, 18-28.
\refs Mandell A. et al. (2013) \emph{Astrophys. J., 779}, 128-145.
\refs {Marley} M.~S., et al. (1999) \emph{Astrophys. J.}, \emph{513}, 879--893.
\refs {Marley} M.~S., et al. (2007), \textit{Astrophys. J., 655}, 541-549.
\refs{Marley} M.~S. and Sengupta S. (2011), \textit{Astrophys. J., 417}, 2874-2881.   
\refs{Marley} M.~S., et al. (2012), \textit{Astrophys. J., 754}, 135.
\refs Marois C. et al. (2008) \emph{Science, 322}, 1348-1352.
\refs Marois C. et al. (2010) \emph{Science, 468}, 1080-1083. 
\refs Maxted P. F. L. et al. (2012) \emph{Mon. Not. R. Astron. Soc., 428}, 2645-2660.
\refs {Menou} K. (2012) \emph{Astrophys. J.}, \emph{744}, L16.
\refs{Metchev} S.~A. and {Hillenbrand}, L.~A. (2006) \emph{Astrophys. J.} \emph{651} 1166.
\refs Miller-Ricci E. et al. (2008) \emph{Astrophys. J., 690}, 1056-1067.
\refs {Miller-Ricci} E., {Seager} S., and {Sasselov} D. (2009) \emph{Astrophys. J.}, \emph{690}, 1056--1067.
\refs {Miller-Ricci} E. and {Fortney} J.~J. (2010) \emph{Astrophys. J.}, \emph{716}, L74--L79.
\refs {Mohanty} S. et al. (2007) \emph{Astrophys. J.}, \emph{657} 1064.
\refs {Morley} C.~V., et al. (2013) \emph{Astrophys. J., 775}, 33-45.
\refs Morris B. M. et al. (2013) \emph{Astrophys. J., 764}, L22-L27.
\refs Moses J. et al. (2011) \textit{Astrophys. J., 737}, 15-51.
\refs Moses J. et al. (2013) \textit{Astrophys. J., 777}, 34-56. 
\refs Mousis et al. (2009)  \textit{Astron. J., 507}, 1671-1674.   
\refs {Mugrauer} M., et al. (2006) \emph{Mon. Not. R. Astron. Soc.}, \emph{373}, L31-L35.
\refs Murgas F. et al. (2012) \emph{Astron. and Astrophys.}, \emph{544}, A41, 5pp.
\refs Murray-Clay, R. A. et al. (2009) \emph{Astrophys. J., 693}, 23-42.
\refs {Nakajima} T. et al. (1995) \emph{Nature}, \emph{378}, 463-465.
\refs {Nettelmann} N. et al. (2011) \emph{Astrophys. J.}, \emph{733}, 2.
\refs {Nielsen} E. L. et al. (2010) \emph{EAS Pub. Series}, \emph{41}, 107-110.
\refs Noll K. S., Geballe T. R., and Marley M. S. (1997) \textit{Astrophys. J., 489}, L87-L90.
\refs Nutzman P. and Charbonneau, D. (2008) \emph{Publ. Astron. Soc. Pacific, 120}, 317-327.
\refs \"{O}berg K., Murray-Clay R., and Bergin E. A. (2011) \emph{Astrophys. J. Lett.}, \emph{743}, L16, 5 pp.
\refs {Oppenheimer} B.~R., et al.  (2013) \emph{Astrophys. J.}, \emph{768}, 24.
\refs {Parmentier} V., {Showman} A., and {Lian} Y. (2013) \emph{Astrophys. J., 558}, A91, 21pp.
\refs {Patience} J. et al. (2010) \emph{Astron. and Astrophys.}, \emph{517}, 76.
\refs Pont F. et al. (2008) \emph{Mon. Not. R. Astron. Soc. 385}, 109-118.
\refs Pont F. et al. (2013) \emph{Mon. Not. R. Astron. Soc. 432}, 2917-2944.
\refs Prinn R. G. and Barshay S. S. (1977) \textit{Science, 198}, 1031.
\refs Rauscher E. and Menou C. (2012) \emph{Astrophys. J., 750}, 96, 13pp.
\refs Redfield S. et al. (2008) \emph{Astrophys. J., 673}, L87-L90.
\refs Richardson L. J. et al. (2007) \emph{Nature, 445}, 892-895.
\refs {Rodler F., L\'opez-Morales M., and Ribas I. (2012) \emph{Astrophys. J.}, \emph{753}, L25-L29.
\refs {Rodler} et al. (2013) \emph{Mon. Not. R. Astron. Soc. 385}, \emph{432}, 1980--1988.
\refs Rodler F. and L\'opez-Morales M. (2014) \emph{Astrophys. J.}, 781, 54, 12pp
\refs {Rogers} L.~A. and {Seager} S. (2010) \emph{Astrophys. J.}, \emph{716}, 1208--1216.
\refs Rothman L. S., et al. (2010) \textit{J. Quant. Spec. \& Rad. Transfer, 111}, 2139-2150.
\refs {Rowe} J.~F., et al. (2008) \emph{Astrophys. J.}, \emph{689}, 1345--1353.
\refs Sanchis-Ojeda R. et al. (2013) \emph{Astrophys. J.}, 774, 54, 9pp
\refs Seager S. and Sasselov D. (1998) \emph{Astrophys. J., 502}, L157-L161.
\refs Seager S. and Sasselov D. (2000) \emph{Astrophys. J., 537}, 916-921.
\refs {Seager} S., {Whitney} B.~A., and {Sasselov} D.~D. (2000) \emph{Astrophys. J.},
  \emph{540}, 504--520.
\refs Seager S., et al. (2005) \textit{Astrophys. J., 632}, 1122-1131.
\refs Seager S., et al. (2007) \textit{Astrophys. J., 669}, 1279-1297.
\refs Seager S. et al. (2009) \emph{Astrophys. and Space Sci. Proc.}, Astrophysics in the Next Decade, eds. H. A. Thronson, et al., Springer Netherlands, 123-145.
\refs Seager S. and Deming D. (2010) \textit{Ann. Rev. Astron. Astrophys., 48}, 631-672.
\refs Shabram M., et al. (2011)  \textit{Astrophys. J., 727}, 65-74.
\refs Sharp C. M. and Huebner W. F. (1990) \textit{Astrophys. J. Supp., 72}, 417-431.
\refs Sharp C. M. and Burrows, A. (2007) \textit{Astrophys. J. Supp., 168}, 140-166.
\refs {Showman} A.~P. and {Guillot} T. (2002) \emph{Astron. and Astrophys.}, \emph{385}, 166--180.
\refs {Showman} A.~P., {Menou} K., and {Cho} J.~Y.-K. (2008) In \emph{Extreme Solar
  Systems} (D. {Fischer} et al., eds.), \emph{Astron. Soc. Pac. Conf. Series}, v. 398, p. 419.
\refs {Showman} A.~P., et al. (2009) \emph{Astrophys. J.}, \emph{699}, 564--584.
\refs Showman A. P. et al. (2009) in \emph{Exoplanets}, ed. S. Seager, U. Arizona Press: Tucson, p. 471-516.
\refs Showman A. P. and Polvani, L. M. (2011) \emph{Astrophys. J., 738}, 71-95.
\refs Showman A. P. et al. (2012) \emph{Astrophys. J., 762}, 24-44.
\refs Showman A. P. and Kaspi Y. (2013) \emph{Astrophys. J.}, 776, 85, 19pp
\refs Showman A. P. et al. (2013) in \textit{Comparative Climatology of Terrestrial Planets, in press}, eds. S. Mackwell, M. Bullock, and J. Harder,  Arizona Space Science Series. 
\refs Sing D. K. et al., (2008) \emph{Astrophys. J., 686}, 658-666.
\refs Sing D. K. and L\'opez-Morales M. (2009) \emph{Astron. and Astrophys., 493}, L31-L34
\refs Sing D. K. et al., (2011) \emph{Mon. Not. R. Astron. Soc. 416}, 1443-1455.
\refs {Sing} D.~K. et al., (2011) \emph{Astron. and Astrophys.}, \emph{527}, A73.
\refs {Skemer} A.~J., et al. (2012) \emph{Astrophys. J.}, \emph{753}, 14.
\refs Snellen I. et al. (2009) \emph{Nature, 459}, 543-545.
\refs Snellen I. et al. (2010a) \emph{Astron. Astrophys., 513}, A76-A85.
\refs Snellen I. et al. (2010b) \emph{Nature, 465}, 1049-1053.
\refs Snellen I. et al. (2013) \textit{Astrophys. J., 764}, 182-187. 
\refs Sozzetti, A. et al. (2013) in \emph{Hot Planets and Cool Stars}, eds. R. Saglia, EPJ Web of Conferences, 47, 3006-3014.
\refs Spiegel D. S., Silverio K., and Burrows A. (2009) \textit{Astrophys. J., 699}, 1487-1500.
\refs Spiegel D. S., et al. (2010). \textit{Astrophys. J., 709}, 149-158.
\refs Spiegel D. S. and Burrows A. (2012) \textit{Astrophys. J., 745}, 174-188.
\refs Stam D. M., Hovenier J. W., and Waters L. B. F. M. (2004) \emph{Astron. and Astrophys., 428}, 663-672.
\refs {Stevenson} D. (1991) \emph{Ann. Rev. Astron. Astrophys.}, \emph{29}, 163-193.
\refs Stevenson K. B. et al. (2010) \emph{Nature}, 464, 1161-1164.
\refs {Sudarsky} D., {Burrows} A., and {Pinto} P. (2000) \emph{Astrophys. J.}, \emph{538}, 885--903.
\refs Sudarsky D. et al. (2003) \emph{Astrophys. J., 588}, 1121-1148.
\refs Swain M. R. et al. (2008a) \emph{Astrophys. J., 674}, 482-497.
\refs Swain M. R. et al. (2008b) \emph{Nature, 452}, 329-331.
\refs Swain M. R. et al. (2009a) \emph{Astrophys. J., 690}, L114-L117.
\refs Swain M. R. et al. (2009b) \emph{Astrophys. J., 704}, 1616-1621.
\refs Swain M. R. et al. (2010) \emph{Nature, 463}, 637-639.
\refs Swain M. R. et al. (2012) \emph{Icarus 225}, 432-445.
\refs Tamura M. (2009) \emph{in AIP Conf. Ser. 1158, ed. T. Usuda, M. Tamura, \& M. Ishii (Melville,
NY: AIP)}, 11 
\refs {Tennyson} J. and {Yurchenko} S.~N. (2012) \emph{Mon. Not. R. Astron. Soc.}, \emph{425}, 21-33.
\refs Tinetti G., et al. (2007) \textit{Nature, 448}, 169-171. 
\refs Tinetti G., et al. (2010)  \textit{Astrophys. J. Lett, 712}, 139, 4pp.
\refs Tinetti G., et al. (2013)  \textit{Experimental Astron., 34}, 311-353.
\refs {Valencia} D., et al. (2013) \textit{Astrophys. J., 775}, 10-21. 
\refs Venot O., et al. (2012) \emph{Astron. Astrophys., 546}, A43-A61.
\refs {Vidal-Madjar} A., et al. (2003) \emph{Nature}, \emph{422}, 143-146.
\refs {Vidal-Madjar} A., et al. (2004) \emph{Astrophys. J.}, \emph{604}, L69-L72.
\refs Visscher C. and Moses J. (2011) \textit{Astrophys. J., 738}, 72-83.
\refs  Waldmann I. P. et al. (2012) \emph{Astrophys. J., 744}, 35-45.
\refs Waldmann I. P. et al. (2013) \emph{Astrophys. J., 766}, 7-16.
\refs Wang W. et al. (2013) \emph{Astrophys. J., 770}, 70, 8pp
\refs Wiktorowicz S. (2009) \emph{Astrophys. J., 696}, 1116-1124. 
\refs Williams P., et al. (2006) \emph{Astrophys. J., 649}, 1020-1027
\refs Winn J. N. et al. (2011) \emph{Astrophys. J., 737}, L18-L24.
\refs Wood P. L. et al. (2011) \emph{Mon. Not. R. Astron. Soc., 412}, 2376-2382.
\refs {Yelle} R., {Lammer} H., and {Ip} W.-H. (2008) \emph{Space Sci. Rev.}, \emph{139}, 437--451.
\refs Yung Y. and Demore W. B. (1999) \textit{Photochemistry of Planetary Atmospheres}, Oxford Univ. Press., NY.
\refs Zahnle K., et al. (2009) \textit{Astrophys. J., 701}, L20-L24. 
\end{document}